\documentclass[twocolumn,times,tighten]{aastex63}
\usepackage{color}
\usepackage{enumitem}
\usepackage{hyperref}
\usepackage{verbatim}
\usepackage{amsmath,amstext,mathrsfs}
\usepackage[all]{hypcap} 
\usepackage{afterpage}
\usepackage{color}
\usepackage{soul}

\newcommand*{\torefereetwo}{}
\newcommand*{\torefereeone}{}

\accepted{by ApJ on May 12, 2020}

\begin{document}
\title{Advancing the Velocity Gradient Technique: Using Gradient Amplitudes and handling thermal broadening}

\author[0000-0003-1683-9153]{Ka Ho Yuen}
\affiliation{Department of Astronomy, University of Wisconsin-Madison}
\email{kyuen@astro.wisc.edu}

\author[0000-0002-7336-6674]{Alex Lazarian}
\affiliation{Department of Astronomy, University of Wisconsin-Madison}
\affiliation{Korea Astronomy and Space Science Institute, Daejeon 34055, Republic of Korea}
\email{alazarian@facstaff.wisc.edu}

\begin{abstract}
The recent development of the Velocity Gradient Technique allows observers to map magnetic field orientations and magnetization using the direction of velocity gradients. Aside from the directions, amplitudes of velocity gradients also contain valuable information {about the underlying properties of magneto-hydrodynamic (MHD) turbulence. In this paper}, we explore what physical information is contained in the amplitudes of velocity gradients and discuss how this information can be used {\torefereeone to diagnose properties of} turbulence in both diffuse and self-gravitating interstellar media. We identify the relations between amplitudes of both intensity and velocity centroid gradients and the sonic Mach number $M_s$ and they are consistent with the {\torefereeone theory's predictions}. We test the robustness of the method {\torefereeone and} discuss how to utilize the amplitudes of gradients into self-gravitating media. To extend the velocity gradient technique we also discuss the usage of amplitude method to Position-Position Velocity (PPV) space as a possible way to retrieve the velocity channel maps before the contamination of thermal broadening. We discuss that the Velocity Gradient Technique with these advancements could potentially give a significantly more accurate statistical insight into the properties magnetized turbulence.
\end{abstract}

\section{Introduction}

Turbulence is ubiquitous in astrophysical environment. Magneto-hydrodynamic (MHD) turbulence plays a very important role in various astrophysical phenomena (see Larson 1981,  \citealt{Armstrong1995ElectronMedium, Chepurnov2010ExtendingData,2003matu.book.....B,2004ARA&A..42..211E}, Xu \& Zhang 2016), including star formation (see \citealt{MO07,MK04,2016ApJ...824..134F}), propagation and acceleration of cosmic rays (see \citealt{J66,YL08}), as well as regulating heat and mass transport between different ISM phases (\citealt{1993MNRAS.262..327G,  2000ApJ...543..227D,LP04,LP06,2001ApJ...561..264D,2006ApJS..165..512K,2006MNRAS.372L..33B,2006ApJ...653L.125P} see \citealt{Draine2009} for the list of the phases). Our understanding of MHD turbulence has been significantly improved due to {\torefereeone the establishment of the anisotropic scaling law} \citep{GS95}, quantitative description of turbulent reconnection that is a part and parcel of the MHD cascade \citep{LV99} as well as numerical studies that tested {the} theoretical ideas \citep{CV00, MG01, CL02,CL03,KLV09,2010ApJ...720..742K}. A detailed discussion of the present understanding of MHD turbulence can be found in \cite{2019tuma.book.....B}.

Supported by the advancement of {\torefereeone the} MHD turbulence theory, the idea of using spectroscopic velocity gradients to trace magnetic field, namely the Velocity Gradient Technique (VGT, \citealt{GL17,YL17a,YL17b,LY18a}), was recently proposed and applied to observation data including galactic HI and self-gravitating molecular clouds. The recently development of VGT allows the measurements of magnetic field orientations by probing the peak of the velocity gradients orientation histogram in a statistically well-sampled area \citep{YL17a}. The orientations of velocity gradients are taken into account for tracing both the direction of magnetic field in diffuse \citep{YL17a} and self-gravitating media \citep{YL17b,LY18a} based on turbulence scaling with the concept of block-averaging \citep{YL17a}. The VGT has demonstrated its capability to trace magnetic fields in different interstellar (ISM) phases. Similar techniques based on the structures of turbulence have also been developed previously \citep{EL05, 2008ApJ...680..420H,2009ApJ...699.1092H}, but the method of gradients was shown to be better in comparison to them \citep{2018ApJ...865...54Y}. In some regimes of turbulence (see \citealt{2005ApJ...624L..93B, 2007ApJ...658..423K}), the turbulence velocity imprints its statistics on density. In this situation, Intensity gradients (IGs) can also be used to trace magnetic field as shown in \cite{LY18a}. The combined used of VCG and IG can also provide extra physical gauge in both self-gravitating and shock regions \citep{YL17b}, and has been applied successfully to observational data \citep{survey,velac}.

We would like to note that the aforementioned IG technique, which is an offshoot of velocity gradient research, should be distinguished from the Histograms of Relative Orientation (HRO) technique (\citealt{Soler2013,SH17},see also \citealt{IGVHRO} for a comparison between IG and HRO). Unlike the IGs under the block-averaging approach in \cite{YL17a}, HRO is not a tool aiming in tracing the directions of magnetic field but is used to study the relative orientation of gradients and polarization statistically as a function of the column density  (See \citealt{IGVHRO}). 

The previous works on gradient technique mostly focus on the direction of the gradients of an observable (Intensity, Centroid, Velocity Channel) within a statistical region of interest. Recently \cite{LYH18} shows that the dispersion of velocity gradients in a sampling block is related to the local magnetization in the diffuse media. The statistics of gradient amplitudes in the framework of VGT and IGs have been addressed in \cite{YL17b} when considering the filtering of shocks in the statistical prediction of gradient orientations. It was also suggested in \cite{LYH18} that amplitudes of gradients in a statistical region of interest could possibly reflect the compressible nature of turbulence. However a comprehensive discussion on the use of the gradient amplitudes similar to that for gradient orientations (e.g \citealt{LY18a}) and dispersions \citep{LYH18} does not exist. 

This paper continues our study of the utility of {\torefereeone gradient method in characterizing the physical} properties of interstellar turbulence. In particular, we focus on the statistical properties of the gradient amplitudes of the observables. In this paper we focus mostly in obtaining estimates of $M_s$ using gradient amplitudes, especially in the case of local interstellar media where the isothermal condition holds. We first discuss the theoretical expectations in \S \ref{sec:theory}, while our numerical setup is described in  \S \ref{sec:method}. We analyze the gradient amplitudes in \ref{sec:analysis} and discuss the robustness of the method in \S \ref{sec:robust}. We discuss one of the very important applications of the amplitude method that would allow removal of the thermal contributions from the observed position-position-velocity cube \S \ref{sec:deconvolution}. We discuss our work in \S \ref{sec:discussion} and summarize our work in \S \ref{sec:conclusion}.

\section{Statistics of gradient amplitudes based on the theory of MHD turbulence}
\label{sec:theory}

In our previous works, we already discussed about how the directions \citep{LY18a} and the dispersions \citep{LYH18} of gradients are related to the theoretical prediction of MHD turbulence anisotropy (\citealt{GS95,LV99,CV00,MG01}, see \citealt{2013SSRv..178..163B} for a review and \citealt{2019tuma.book.....B} for a textbook). Readers can refer to our previous works on how the statistical averaged {parameters} in MHD turbulence are related to both gradients directions and dispersive quantities. Since \cite{GS95} is a statistical theory describing the turbulence motion in the Eulerian frame, the respective analysis in the framework of gradients should also be performed statistically, either by obtaining averaged quantities spatially, temporally or over ensembles. One realization of this statistical averaging is the sub-block averaging proposed in \cite{YL17a}. In this paper, we will mainly focus on the prediction of the statistical properties of gradient amplitude based on the MHD theory of \cite{GS95}. {\torefereeone Note that in the \cite{GS95} framework Alfven modes do not contribute to density fluctuations, which we are discuss in \S \ref{subsec:stat_dens}.}

\subsection{Incompressible limit}
\label{subsec:stat_incompressible}
Based on the theory of {\it incompressible} MHD turbulence suggested by \cite{GS95}, the turbulent eddies are elongated along the magnetic field direction. In the {\torefereeone original} theory by \cite{GS95} it is assumed that the latter direction is to be mean field direction. However, in fact, the turbulent eddies are aligned with the direction of {\torefereeone {\it local} magnetic field}. This follows from the ability of turbulent eddies to perform mixing motions that minimally bend magnetic {\torefereeone fields}. This ability comes from turbulent reconnection  \cite{LV99} that allows {\torefereeone magnetized turbulent} eddies to change the magnetic topology within one eddy turnover time. This dynamics of magnetized eddies in the local reference frame is proved numerically in \cite{CV00,MG01}. The mixing motions of an {\torefereeone eddy} induce {\torefereeone a shearing force maximally} perpendicular to {\torefereeone the eddy's} rotational axis. As this axis coincide with the local magnetic field direction, the {\it $90^o$-rotated} gradient of the absolute value of the eddy velocity {\torefereeone aligns with} the direction of magnetic field at the location of the eddy. 

The \cite{GS95} was formulated assuming that the turbulent injection velocity $V_L$ is equal to the Alfven velocity $V_A$. This means that the Alfvenic Mach number $M_A=V_L/V_A$ was assumed to be 1. Realistic astrophysical settings present a variety of $M_A$ and the theory covering different magnetizations was formulated in \cite{LV99}. Below we reproduce the expressions derived there.  For the case of sub-Alfvenic turbulence \citep{Lazarian2006}, we consider only the strong turbulence with the scale $l=1/k<LM_A^2 = l_{tr}$ since the weak turbulence usually contains very limited spatial range unless $M_A\ll 1$. Then the velocity amplitude $v_l$ at scale $l$ would be:
\begin{equation}
v_l\sim V_L \left(\frac{l_\perp}{L_{inj}}\right)^{1/3} M_A^{1/3},
\label{eq:stg_vscaling}
\end{equation} 
with  $L_{inj}$ being the injection scale, while $l_\perp$ represents the perpendicular (to magnetic field) length scale of the eddies. Since eddies are anisotropic along the magnetic field directions, we expect the gradients of velocities to be in the form $v_l/l_\perp$ as $l_\perp \ll l_\ll$. As a result we would have a prediction of the velocity gradient amplitude {\it of turbulence eddies of scale $l$}:
\begin{equation}
    \nabla v_l = \frac{v_l}{l_\perp} \sim \frac{V_L}{L_{inj}} \left(\frac{L_{inj}}{l_\perp}\right)^{2/3} M_A^{1/3} \quad (M_A<1, l<l_{tr})
    \label{eq:vga_fundamental}
\end{equation}

Similar expressions could be obtained in the case of super-Alfvenic turbulence $M_A>1$ in the incompressible limit. In the case of super-Alfvenic turbulence, there is a transition scale $l_A = LM_A^{-3}$. When the length scale $l>l_A$, then the eddy is isotropic (i.e. $l=l_\perp = l_\parallel$), and the velocity gradient amplitude term would simply be:
\begin{equation}
\nabla v_l \sim  \frac{V_L}{L_{inj}} \left(\frac{L_{inj}}{l}\right)^{2/3} \quad (M_A>1, l>l_A)
\label{eq:superalf1}
\end{equation} 
while in the case of $l<l_A=LM_A^-3$, the anisotropy of eddies still exist and follow the magnetic field line. Hence $l_\perp<l_\parallel$ and 
\begin{equation}
\nabla v_l \sim  \frac{V_L}{L_{inj}} \left(\frac{L_{inj}}{l_\perp}\right)^{2/3} \quad (M_A>1, l<l_A)
\label{eq:superalf2}
\end{equation} 
In simulations, the parameters $V_L$ and $L_{inj}$ are given at the start of the turbulence driving and stay constant under temporal evolution. The three-dimensional velocity gradient amplitude is the sum of velocity gradient amplitudes from eddies of different scales, which we would expect the smallest scale permissible in the simulation dominates. As a result, we would have the simplified, easily memorable expression for the amplitudes of velocity gradients which we would use throughout the whole paper:
\begin{equation}
\nabla v_l \propto {l_\perp}^{-2/3} \min(M_A^{1/3},1)
\label{eq:superalf3}
\end{equation} 

{\torefereeone In incompressible turbulence, there should have no density fluctuations $\delta \rho$ since Alfven modes do not induce such fluctuations. As a result, the density gradients are not considered in incompressible limit of MHD turbulence.}

\subsection{Compressible Turbulence}
\label{subsec:stat_compressible}
In the case of compressible turbulence, three MHD mode arises, namely the incompressible Alfven mode and the two compressible modes called fast and slow modes. For magnetically dominated low-$\beta$ media, fast modes arise from the compression of magnetic field lines in its perpendicular directions while slow modes are compression along the magnetic field lines. In the media dominated by gas pressure, or high-$\beta$ media, the fast modes are similar to sound waves and slow modes are density perturbations propagating along magnetic field directions. \cite{CL02} showed that the driving of compressible modes from Alfven modes is marginal, provided that either the external magnetic field or the gas pressure is sufficiently high. Since the Alfven mode expressions are the same as what is derived above, below we shall follow the framework of \cite{CL03} in deriving the relevant expressions of velocity and density gradient amplitudes for the two compressible MHD modes.

The methodology in deriving the velocity gradient amplitude $\nabla v_l$ for eddies of scale $l$ here follows from the two expressions: (1) How the energy spectrum behaves: $E(k)\propto k^{-p-1}$, or $E(l) \sim v_l^2 \propto l^p \text{ or } l_\perp^p $ for some p, and (2) whether the system is anisotropic $l_\parallel \propto l_\perp^q$ for some q. Then the expected relation for velocity gradient amplitude would be $\nabla v_l \sim l_\perp^{p/2-1}$ if $q<1$ and $\nabla v_l \sim l^{p/2-1}$ otherwise. For instance, if we have the Kolmogorov energy spectrum $E(k)\sim k^{-5/3}$, that would suggests that $v_l^2 =E(l) \sim l^{2/3}$ and thus having the same power law as in Eq.\ref{eq:stg_vscaling}. Below we shall discuss the cases according to the plasma beta $\beta = 2M_A^2/M_S^2$ where $M_S=v_L/c_s$ is the sonic mach number, $c_s$ is the thermal speed. 

If the turbulent flow is highly supersonic, a steeper {\torefereeone velocity} spectral slope of $-2$ is consistently seen in the simulations \citep{F10,2010ApJ...720..742K} for all three modes, suggesting that $\nabla v_{l,alf/slow/fast}\propto l^{-1/2}$ in highly supersonic regime.

For slow modes in high $\beta$ and mildly supersonic low $\beta$ regime ($M_S<=2.3$ as defined in \cite{CL03}, the velocity scaling relation follows completely from the Alfven mode case, as a result the dependence of velocity gradient amplitudes towards the length scale and Mach numbers would follow what we derived from the previous subsection, i.e. $\nabla v_{l,slow}\propto l^{-2/3}$. 

Fast modes do not have {\torefereeone changes} of the spectral behavior with respect to $\beta$ as long as $M_s$ is mildly small. It follows a spectrum of $E_k \sim k^{-3/2}$ and isotropic in small scales $k_\perp=k_\parallel$. Therefore we would expect the velocity gradient amplitude of the fast mode eddies follows $\nabla v_{l,fast}\propto l^{-3/4}$. 

\subsection{Density gradient amplitudes for slow and fast modes}
\label{subsec:stat_dens}
The density of the eddies at Fourier scale $k=1/l$ can also be written as, following \cite{CL03}
\begin{equation}
    |\rho_k| = \frac{\rho_0 v_k}{c} |{\bf \hat{k}}\cdot {\bf \hat{\zeta}}|
    \label{eq:rhok}
\end{equation}
where $\rho_0$ is the mean density, $\hat{\zeta}$ is the unit vector for the respective mode as listed in the Appendix of \cite{CL03} and $c$ is the respective sound speed for the MHD mode. From here we can write the (three dimensional) density gradient amplitude as :
\begin{equation}
    \nabla \left(\frac{|\rho_l|}{\rho_0}\right) \sim \frac{\rho_l}{\rho_0 l} \sim  \nabla v_l \left(c^{-1} \mathscr{F}^{-1}|{\bf \hat{k}}\cdot {\bf \hat{\zeta}}|\right)
    \label{eq:rhok2}
\end{equation}
where we still keep the dot product term in the Fourier space with the inverse Fourier transform operator $\mathscr{F}^{-1}$ to remind the reader that an inverse Fourier Transform should be done after the dot product (\citealt{CL03, 2006ApJ...638..811C}). One can observe that the density gradient amplitude is simply the velocity gradient amplitude multiplied by two extra terms: the dot product term and the sound speed contribution. The dot product term together with the sound speed {\torefereeone have} been derived previously in \cite{CL03}:
\begin{equation}
    \begin{aligned}
    \nabla \left(\frac{|\rho_l|}{\rho_0}\right) &\propto \nabla v_l M_s \quad &(\beta \ll1, \text{slow modes})\\
    \nabla \left(\frac{|\rho_l|}{\rho_0}\right) &\propto \nabla v_l M_A \quad &(\beta \ll1, \text{fast modes})\\   
    \nabla \left(\frac{|\rho_l|}{\rho_0}\right) &\propto \nabla v_l M_s^2/M_A \quad &(\beta \gg1, \text{slow modes})\\    
    \nabla \left(\frac{|\rho_l|}{\rho_0}\right) &\propto \nabla v_l M_s^2 \quad &(\beta \gg1, \text{fast modes})\\
    \end{aligned}
    \label{eq:rhodep}
\end{equation}
The {\torefereeone expectations} of the dependence of mode gradient amplitudes {\torefereeone are} summarized in Table \ref{tab:pred}. The exact relation of how the velocity and density gradient amplitudes would be related to different physical parameters are subject to the ratio of the MHD modes in the environment. \cite{CL02} suggested that Alfven modes are usually predominant in interstellar media and fast modes are infinitesimal suggesting that we could possibly only look at the Alfven mode and slow mode dependence (Table \ref{tab:pred}) when considering the velocity and density gradient amplitudes respectively.

\begin{table*}
\centering
\begin{tabular}{c c c c c c c c}
& \multicolumn{3}{c}{$\beta\ll1, M_s<1$} & \multicolumn{3}{c}{$\beta\gg 1, M_s<1$} & $M_s \gg 1$ \\ 
& Alfven mode & Slow mode & Fast mode & Alfven mode & Slow mode & Fast mode & All MHD modes\\\hline\hline
$\nabla v_l$ & $\propto l^{-2/3}\min(M_A^{1/3},1)$ & $\propto  l^{-2/3}$ & $\propto l^{-3/4}$ & $\propto l^{-2/3}\min(M_A^{1/3},1)$ & $\propto l^{-2/3}$ & $\propto l^{-3/4}$ & $\propto l^{-1/2}$\\
$\nabla \left(\frac{|\rho_l|}{\rho_0}\right)$ &N/A&$\propto l^{-2/3} M_s$&$\propto l^{-3/4}M_A$&N/A&$\propto l^{-2/3}M_s^2/M_A$&$\propto l^{-3/4}M_s^2$&Follow Eq.\ref{eq:rhodep}\\\hline \hline
\end{tabular}
\caption{\label{tab:pred} Predictions on the dependence of gradient amplitudes as a function of length scale $l$, sonic Mach number $M_s$ and Alfvenic Mach Number $M_A$ for different regimes.}
\end{table*}

\subsection{Observational diagnostics for gradient amplitudes}
\label{subsec:stat_observation}
Since we do not observe the three dimensional velocity and density gradient amplitudes in observation, it is necessary to discuss how the observables, namely the column density (intensity) $I = \int \rho dz$ ,velocity centroid $C=\int \rho v_z dz/I$ and velocity channel maps are expected to behave based on the discussion in the previous subsections. In the following we shall discuss first the intensity and centroid maps and postpone the discussion of channel map gradients to later sections.

\subsubsection{{\torefereeone Relationship} between the two dimensional and three dimensional gradient amplitude quantities}

We first assume the mean magnetic field is on the plane of sky and take the intensity gradient amplitude as an example. The quantity $\nabla_{2D} I = \int \nabla_{2D}\rho dz$ suggests that (1) the semi-minor axis $\l_\perp$ lines on the plane of sky suggests that $\nabla_{2D}\rho = \nabla \rho$ (2) the sign of the density gradient amplitudes along the line of sight follows a random walk. Following a similar argument as in Cho \& Yoo (2016), assuming we have $N=L_{los}/L_{inj}$ eddies along the line of sight in the stage of the injection, then we would have an estimation of intensity gradient amplitudes
\begin{equation}
    \nabla_{2D} I \propto \nabla \rho L_{inj} \sqrt{N}
    \label{eq:Irho}
\end{equation}
which suggests that the intensity gradient amplitude is proportional to the density gradient amplitudes with extra factors that would be a constant in time in simulations. If the mean magnetic field makes an angle of $\theta$ to the line of sight magnetic field, then there would be an extra $\cos\theta$ factor in Eq.\ref{eq:Irho} due to the conversion of $\nabla_{2D}$ to $\nabla_{3D}$, i.e. $\nabla_{2D} I \propto \nabla \rho \cos\theta$. {\torefereeone However,as shown in the appendix of Lazarian et.al 2020 (See also \citealt{BLOS}), the insertion of the $\cos\theta$ in compensating the effect of line-of-sight angle is not correct. We can understand the argument in Lazarian et.al (2020) by considering a magnetic field line with an inclination angle $\theta$. Conceptually, the magnetic field contains both mean and the turbulent components, which we can assume the latter is perpendicular to the mean field. In the case when the mean field component is almost aligned with the line of sight, the observed magnetic field would be contributed mostly the turbulent component of the magnetic field plus a small contribution from the mean field. If it happens that the Alfvenic Mach number is large enough, i.e. the turbulent-to-mean magnetic field strength ratio is large, then the observed magnetic field on the plane of sky is {\it larger} than the expected strength computed by $B\cos\theta$. The same argument applies to all vector quantities, in particular, the intensity gradient amplitude as in Eq.\ref{eq:Irho}.From a similar argument in Lazarian et.al (2020), we see that such an approximation formula in Eq.\ref{eq:Irho} is correct only when the $\theta> 4\tan^{-1}(M_A/\sqrt{3})$.}

The situation in velocity centroid is more complicated since there are two terms upon differentiation. \cite{EL05} showed that the velocity centroid is a proper measure of velocities. However upon planar differentiation how would the density and velocity gradient term contribute relatively is unknown and requires more substantial research. From previous numerical work on VGT \citep{LY18a}, it is apparent that the velocity weighting in the calculation of velocity centroids increases the velocity nature in terms of the structures, regardless of the normalization constant. Therefore we would expect the velocity centroid to be "more velocity like" (See \citealt{Kandel17}). 

Velocity channels \citep{LP00,LP04} are even more complicated than velocity centroids due to (1) the decreased velocity channel width would then increases the velocity weighting in the channel map (2) the increase of thermal broadening washes away the velocity nature in the channel map. Both factors depend on the {\torefereeone spectra} of densities and velocities, as well as the sonic Mach number of the interstellar medium. It is expected that in the case of high sonic Mach number, the shocks together with the velocity ambient map would be more apparent in the velocity channel map \citep{YL17a,LY18a}. A suitable thermal deconvolution algorithm is required if sonic Mach number is too low (See \S \ref{sec:deconvolution}).

\subsubsection{Proper Statistical averaging in observations}

As discussed above, some sort of statistical averaging {\torefereeone is} required to make the prediction of MHD theory (Table \ref{tab:pred}) applicable. A straightforward statistical measure would be to compute the mean and dispersion of the gradient amplitudes within a large enough statistical sampling area. We take the case of velocity gradient amplitude as an example since the respective observables are directly proportional to the velocity gradient amplitudes as shown in previous subsections. From Table \ref{tab:pred} we can safely assume $\nabla v \propto l^a M_A^b$ for some number $a,b$. Given an averaging volume $V$, the sonic and Alfvenic Mach numbers are computed already. The mean and dispersion of velocity gradient amplitude {\torefereeone are} simply the mean and the dispersion of the respective length scale mean and dispersion times the relevant Mach numbers, i.e.
\begin{equation}
\begin{aligned}
    \langle \nabla v_l\rangle_l \propto \langle l^a \rangle_l M_A^b
    \sigma_{\nabla v_l} &\sim \sigma_{l^a} M_A^b
\end{aligned}
\end{equation}
therefore using these simple statistical quantities would allow observers to trace the sonic and Alfvenic pre-factor respectively.

As a summary of the current section, we derive the incompressible three-dimensional velocity gradient amplitude relation based on the MHD theory (\S \ref{subsec:stat_incompressible}). Based on that we further proceed in deriving the velocity gradient amplitude relation for different modes in compressible turbulence (\S \ref{subsec:stat_compressible}), which we found that the Alfven mode would be the dominant term in the gradient amplitudes of velocities. The respective density gradient amplitude relations for slow and fast modes are derived in (\S \ref{subsec:stat_dens}) and we predict that the slow modes would then dominate the fluctuation of density gradient amplitudes. In \S \ref{subsec:stat_observation} we{\torefereeone are}onnect the gradient amplitudes of two-dimensional observables to the three dimensional counterparts and further show that the simple statistical measures of intensity and velocity gradient amplitudes would allow observers to measure the Mach Numbers as listed in Table \ref{tab:pred}. From our deviation in Table \ref{tab:pred}, it is apparently easier to explore the relation between the gradient amplitude of the observables to $M_s$ since the dependence of the gradient amplitude of the observables on $M_A$ is highly nonlinear and mode dependent. 

\begin{figure*}[t]
\centering
\includegraphics[width=0.98\textwidth]{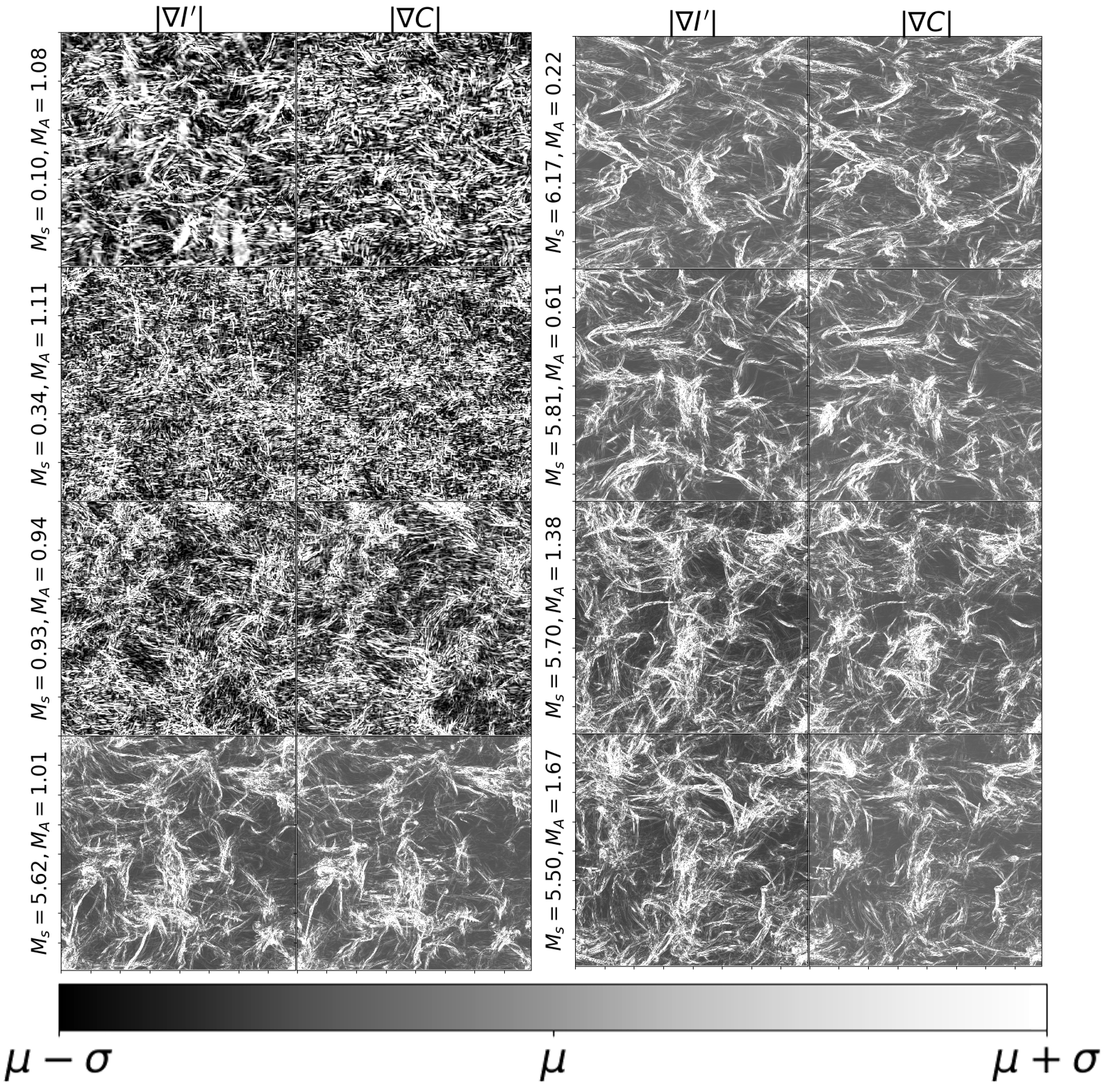}
\caption{\label{fig:illus} Visual {\torefereeone illustrations} of how $M_s$ (left column) and $M_A$ (right column) changes gradients of intensities and gradients of centroids, i.e. $|\nabla I|$, $|\nabla C|$, that we study here. The colorbar scale are the same for all plots, that the darkest color corresponds to the mean value minus the standard variation of the respective map, while the lightest color corresponds to the mean value plus the standard variation. Notice the last row }
\end{figure*}
\section{Method}
\label{sec:method}
The numerical data cubes {\torefereeone are} obtained by 3D MHD simulations that {\torefereeone are} from a single fluid, operator-split, staggered grid MHD Eulerian code ZEUS-MP/3D to set up a three dimensional, uniform turbulent medium. {\torefereeone Most of our simulations are isothermal with $T=10K$ but some of them are adiabatic by assuming the $P\propto \rho^{5/3}$ to mimic the properties of the warm neutral media \citep{2017NJPh...19f5003K}.} To simulate the part of the interstellar cloud, periodic boundary conditions are applied. We inject turbulence solenoidally\footnote{These simulations are the Fourier-space forced driving isothermal simulations. {The choice of force stirring over the other popular choice of decaying turbulence is because only the former will exhibit the full characteristics of turbulence statistics (e.g power law, turbulence anisotropy) extended from $k=2$ to {\torefereeone a dissipation scale of $12$ pixels} in a simulation , and matches with what we see in observations (e.g. \citealt{Armstrong1995ElectronMedium, Chepurnov2010ExtendingData}) }.} 

For our controlling simulations parameters, various Alfvenic Mach numbers $M_A=V_{inj}/V_A$ and sonic Mach numbers $M_s=V_{inj}/V_s$ are employed \footnote{ For isothermal MHD simulation without gravity, the simulations are scale-free. The two scale-free parameters $M_A,M_s$ determine all properties of the numerical cubes and the resultant simulation is universal in the inertial range. That means one can easily transform to whatever units as long as the dimensionless parameters $M_A,M_s$ are not changed.}, where $V_{inj}$ is the injection velocity, while $V_A$ and $V_s$ are the Alfven and sonic velocities respectively, which they are listed in Table \ref{tab:sim}. For the case of $M_A<M_s$, it corresponds to the simulations of turbulent plasma with thermal pressure smaller than the magnetic pressure, i.e. plasma with low confinement coefficient $\beta/2=V_s^2/V_A^2<1$. In contrast, the case that is $M_A>M_s$ corresponds to the magnetic pressure dominated plasma with high confinement coefficient $\beta/2>1$. 

Further we refer to the simulations in Table \ref{tab:sim} by their model name. For example, the figures with model name indicate which data cube was used to plot the corresponding figure. Each simulation name follows the rule that is the name is with respect to the varied $M_s$ \& $M_A$ in ascending order of confinement coefficient $\beta$. The selected ranges of $M_s, M_A, \beta$ are determined by possible scenarios of astrophysical turbulence from very subsonic to supersonic cases. 

The raw data from simulation cubes are converted to synthetic maps for our gradient studies. The normalized velocity centroid $C({\bf R})$ in the simplest case\footnote{Higher order centroids are considered in \cite{YL17b} and they have $v^n$, e.g. with $n=2$, in the expression of the centroid. Such centroids may have their own advantages. However, for the sake of simplicity we employ for the rest of the paper $n=1$.} {\torefereeone are} defined as
\begin{equation}
\begin{aligned}
C({\bf R}) &=I^{-1} \int \rho_v({\bf R},v) v  dv,\\
I({\bf R}) &= \int \rho_v({\bf R},v)  dv,
\end{aligned}
\label{centroid}
\end{equation}
where $\rho_v$ is density of the emitters in the Position-Position-Velocity (PPV) space, $v$ is the velocity component along the line of sight and ${\bf R}$ is the 2D vector in the pictorial plane. The integration is assumed to be over the entire range of $v$.  Naturally, $I({\bf R})$ is the emission intensity. The $C(R)$ is also an integral of the product of velocity and line of sight density, which follows from a simple transformation of variables (see \citealt{2003ApJ...592L..37L}). For constant density, $C(R)$ is just a velocity averaged over the line of sight. 

\begin{table}
\centering
\begin{tabular}{c c c c c}
Model & $M_S$ & $M_A$ & $\beta=2M_A^2/M_S^2$ & Resolution \\ \hline \hline
m0                      & 5.73  & 0.22 & 0.0029 & $360^3$ \\
m1                      & 5.79  & 0.42 & 0.011 & $360^3$ \\
m2                      & 5.69  & 0.61 & 0.023 & $360^3$ \\
m3                      & 5.46  & 0.82 & 0.045 & $360^3$ \\
m4                      & 5.50  & 1.01 & 0.067 & $360^3$ \\
m5                      & 5.51  & 1.19 & 0.093 & $360^3$ \\
m6                      & 5.45  & 1.38 & 0.13 & $360^3$ \\
m7                      & 5.41  & 1.55 & 0.16 & $360^3$ \\
m8                      & 5.41  & 1.67 & 0.19 & $360^3$ \\
m9                      & 5.40  & 1.71 & 0.20 & $360^3$ \\ \hline
Ms0.4Ma0.04             & 0.41  & 0.04 & 0.02 & $480^3$ \\
Ms0.8Ma0.08             & 0.92  & 0.09 & 0.02 & $480^3$ \\
Ms1.6Ma0.16             & 1.95  & 0.18 & 0.02 & $480^3$ \\
Ms3.2Ma0.32             & 3.88  & 0.35 & 0.02 & $480^3$ \\
Ms6.4Ma0.64             & 7.14  & 0.66 & 0.02 & $480^3$ \\ 
Ms0.4Ma0.132            & 0.47  & 0.15 & 0.22 & $480^3$ \\
Ms0.8Ma0.264            & 0.98  & 0.32 & 0.22 & $480^3$ \\
Ms1.6Ma0.528            & 1.92  & 0.59 & 0.22 & $480^3$ \\ 
Ms0.4Ma0.4              & 0.48  & 0.48 & 2.0 & $480^3$ \\
Ms0.8Ma0.8              & 0.93  & 0.94 & 2.0 & $480^3$ \\ 
Ms0.132Ma0.4            & 0.16  & 0.49 & 18 & $480^3$ \\
Ms0.264Ma0.8            & 0.34  & 1.11 & 18 & $480^3$ \\ 
Ms0.04Ma0.4             & 0.05  & 0.52 & 200 & $480^3$ \\
Ms0.08Ma0.8             & 0.10  & 1.08 & 200 & $480^3$ \\ \hline
huge-0                  & 6.17  & 0.22 & 0.0025 & $792^3$ \\
huge-1                  & 5.65  & 0.42 & 0.011 & $792^3$ \\
huge-2                  & 5.81  & 0.61 & 0.022 & $792^3$ \\
huge-3                  & 5.66  & 0.82 & 0.042 & $792^3$ \\
huge-4                  & 5.62  & 1.01 & 0.065 & $792^3$ \\
huge-5                  & 5.63  & 1.19 & 0.089 & $792^3$ \\
huge-6                  & 5.70  & 1.38 & 0.12 & $792^3$ \\
huge-7                  & 5.56  & 1.55 & 0.16 & $792^3$ \\
huge-8                  & 5.50  & 1.67 & 0.18 & $792^3$ \\
huge-9                  & 5.39  & 1.71 & 0.20 & $792^3$ \\ \hline
h0-1200                 & 6.36  & 0.22 & 0.00049 & $1200^3$ \\
h9-1200                 & 10.79 & 1.26 & 0.0068 & $1200^3$ \\
e5r2                    & 0.13  & 1.57 & 292 & $1200^3$ \\
e5r3                    & 0.61  & 0.52 & 1.45 & $1200^3$ \\
e6r3                    & 5.45  & 0.24 & 0.0039 & $1200^3$ \\
e7r3                    & 0.53  & 1.31 & 12.22 & $1200^3$ \\
h0-1600                 & 5.56  & 0.20 & 0.0026 & $1600^3$ \\ \hline
Ms0.2Ma0.2              & 0.2   & 0.2  & 2 & $480^3$ \\
Ms0.4Ma0.2              & 0.4   & 0.2  & 0.5 & $480^3$ \\ 
Ms4.0Ma0.2              & 4.0   & 0.2  & 0.005 & $480^3$ \\
Ms20.0Ma0.2             & 20.0  & 0.2  & 0.0002 & $480^3$ \\ \hline \hline
\end{tabular}
\caption{\label{tab:sim} Description of MHD simulation cubes {which some of them have been used in the series of papers about VGT \citep{YL17a,YL17b,LY18a,2018ApJ...865...59L}}.  $M_s$ and $M_A$ are the R.M.S values at each the snapshots are taken. }
\end{table}

\section{Gradient amplitude statistics in numerical simulations}
\label{sec:analysis}

Driven by the theoretical derivation from \S \ref{subsec:stat_dens}, it is very natural to consider the quantity $\nabla I'=\nabla (I/\langle I\rangle)$ since the corresponding density gradient amplitude is in the form $\nabla (\rho/\langle \rho \rangle)$, which is obtained by the Sobel kernel. The mean $\mu$ and dispersion $\sigma$ is then defined as:
\begin{equation}
    \begin{aligned}
    \mu &= \langle |\nabla I'|\rangle\\
    \sigma &= \langle |\nabla I'|^2\rangle-\mu^2\\
    \end{aligned}
\end{equation}
which we would compute the average over the whole intensity map. {\torefereeone Notice that the gradient amplitudes have an inverse dependence of the length, which in our code unit each $\nabla$ operator multiplies a factor of $n_x/10pc$ where $n_x$ is the resolution of the cubes. }

We will first illustrate visually how the structures of gradient amplitudes are correlated with the dimensionless physical parameters. Fig \ref{fig:illus} shows how $|\nabla I'|$, $|\nabla C|$ behave when $M_s$ (left column) and $M_A$ (right column)changing by keeping the other parameter approximately constant. The color scale in Fig. \ref{fig:illus} is set so that the darkest color always corresponds to $\mu-\sigma$ and the lightest color always corresponds to $\mu+\sigma$. One could see that the structure of both gradient amplitudes becomes spatially thinner as $M_s$ increases. We would then expect there exists some correlation between $|\nabla I'|$ and $|\nabla C|$ with respect to the sonic Mach number $M_s$. Comparatively, the gradient amplitude {\torefereeone maps} for both intensity and centroid maps {\torefereeone vary} not much in terms of the spatial width of the structures with respect to $M_A$ .

\begin{figure*}[t]   
\centering
\includegraphics[width=0.48\textwidth]{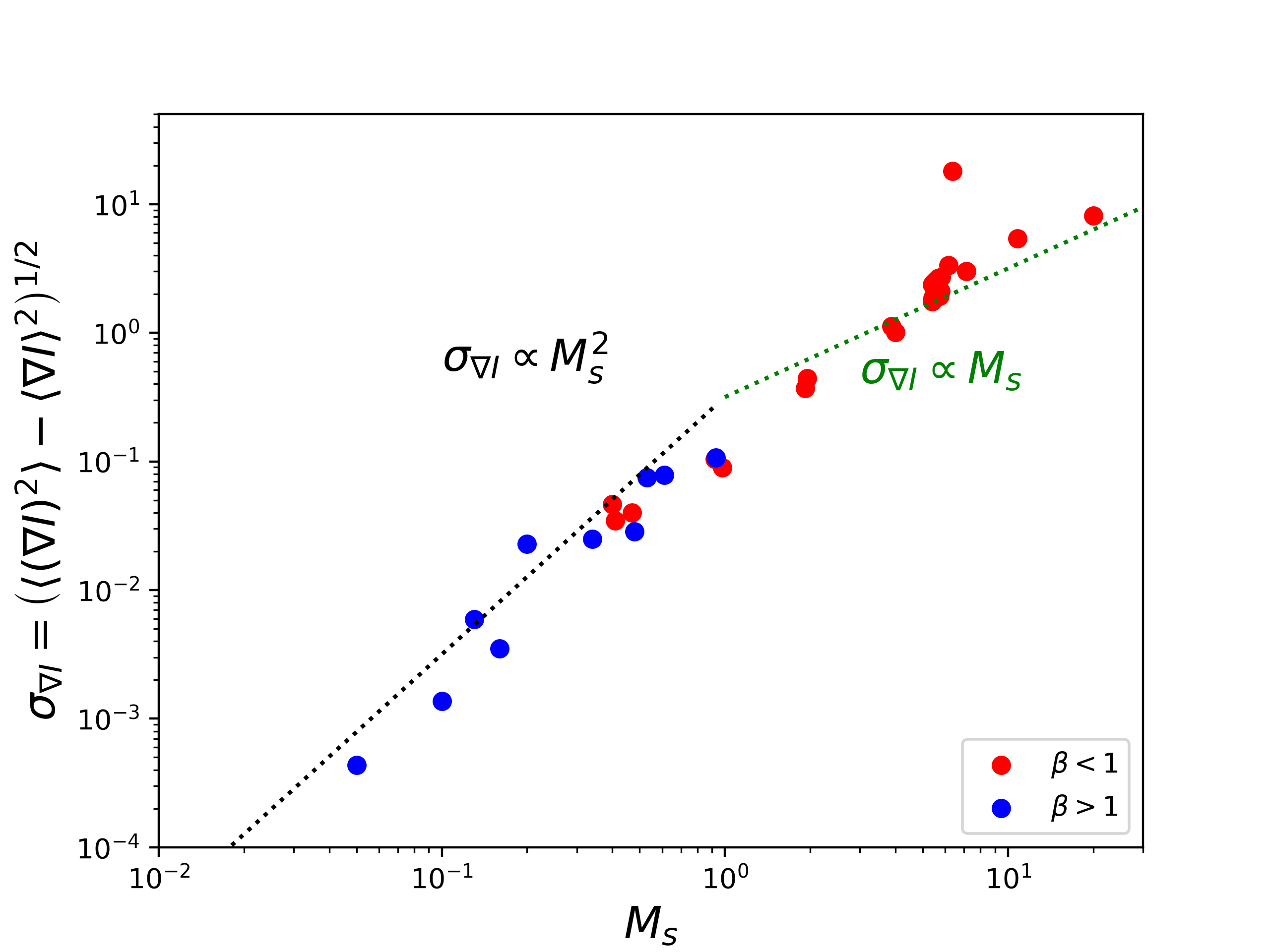}
\includegraphics[width=0.48\textwidth]{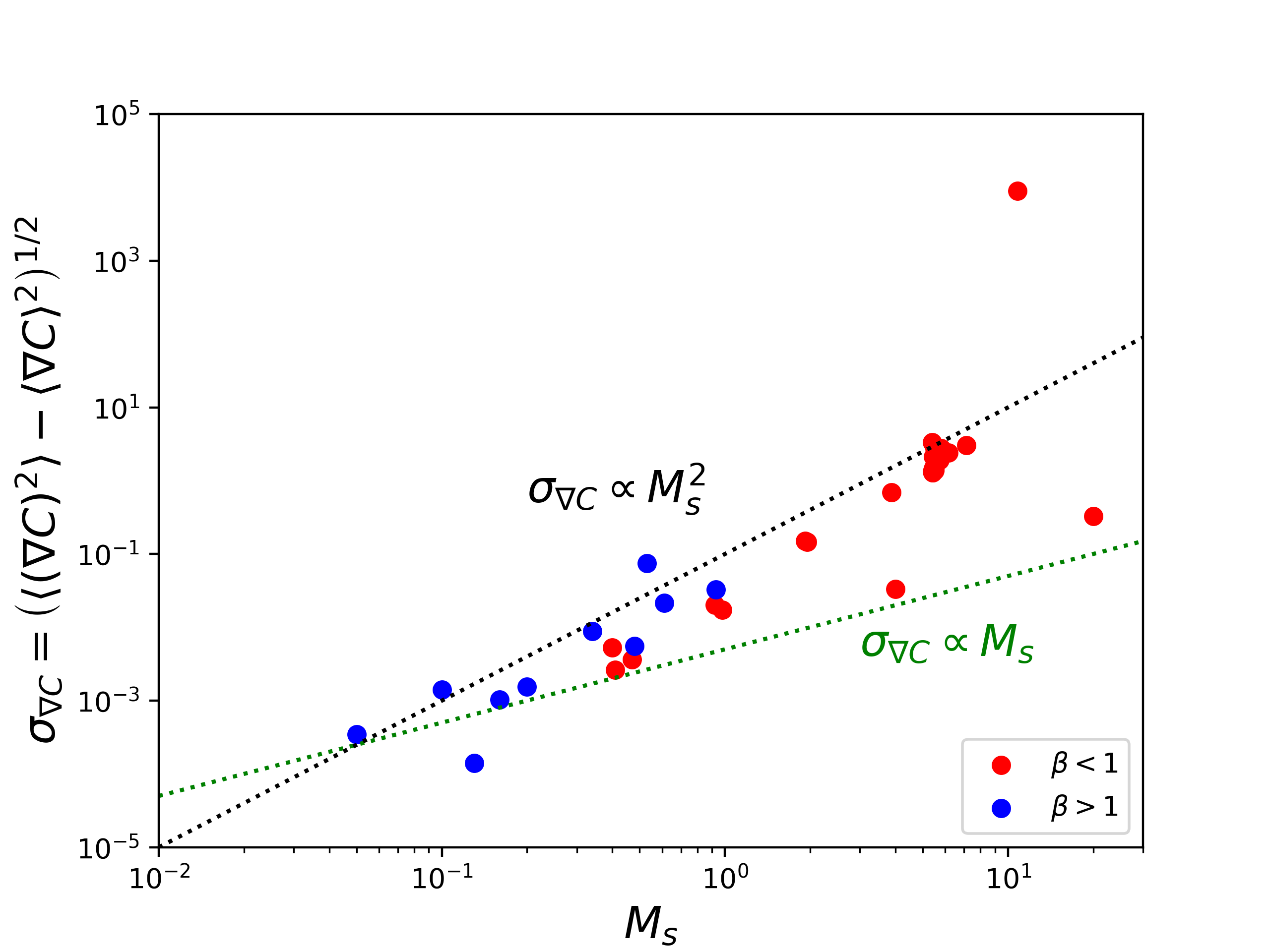}
\caption{\label{fig:n.IGVCGn} A {\torefereeone set of figures} showing the intensity gradient amplitude $\nabla I'=\nabla (I/\langle I\rangle)$ (left) and the centroid gradient amplitude $\nabla C$ (right) as a function of the sonic Mach number $M_s$ ). Due to the prediction in Table \ref{tab:pred}, we denote the data with $\beta>1$ as blue while that with $\beta<1$ as red. We also draw two auxiliary lines (green: $\propto M_s$ and black: $\propto M_s^2$) showing our predicted power laws.}
\end{figure*}

Following the {\torefereeone deviations} from \S \ref{subsec:stat_dens}, we would expect slow mode to dominate the intensity gradient amplitudes. As a result, $\nabla I' \propto M_s$ for $\beta<1$ and $\propto M_s^2$ for $\beta>1$, where we {\torefereeone temporarily} suppress the relation to $M_A$ and the mode weights here. On the left of Fig.\ref{fig:n.IGVCGn} plots the intensity gradient amplitude  $\nabla I'=\nabla (I/\langle I\rangle)$ as a function of the sonic Mach number $M_s$ using the 45 isothermal simulations with different resolutions as we listed in Table \ref{tab:sim}. These data are prepared so that they contain variations of $M_s$ and $M_A$. The {\torefereeone variations are} introduced since we would like to see how strong the one parameter would interfere the power law behavior that we are seeking when we are comparing the $\sigma_{\nabla I'}$ to the other parameter. We draw two auxiliary lines showing our predicted power laws from Table \ref{tab:pred}, one corresponds to the regime where $\nabla I' \propto M_s$ when $\beta\ll1$ while the other {\torefereeone one} corresponds to the regime with $\nabla I' \propto M_s^2$ when $\beta\rightarrow \infty$. We could visually see that the theoretical expectation fairly fits the simulation data. Statistically both the part with $\propto M_s$ and the part with $\propto M_s^2$ have the coefficient of determination to be $\sim 0.89$ and $0.96$ respectively.

Readers should keep in mind that the deviations listed on Table \ref{tab:pred} correspond to the extreme cases of $\beta$ only. It is natural to have different transitional power law when $\beta \sim 1$, as readers might be able to spot that already on the left of Fig. \ref{fig:n.IGVCGn}. In fact, the intermediate regime has not been theoretically studied either in the framework of MHD turbulence, but it has very important astrophysical importance since observationally clouds with e.g. $M_s\sim M_A \sim 1$ are not rare in lukewarm neutral media (despite they have a different thermal properties) or in the case of collapsing core (despite gravity enhances the generation of growing slow and fast modes,). Studies of gradient amplitudes in the intermediate regimes should be combined with the more popular tools like the N-PDF method \citep{2012ApJ...761..149K,BL12} for a better accuracy .

The theory also expects {\torefereeone variations} due to the different {\torefereeone weighting of MHD} modes as $\beta$ changes. For instance, in the case of $\beta\ll 1$ and $M_s$ small, slow modes would contribute to the 3D density gradient amplitude as $\propto l^{-2/3}M_s$ while that of fast modes as $\propto l^{-3/4}M_A$. Similarly in the case of $\beta\gg1$ and $M_s$ small, the corresponding power law is $\propto l^{-2/3}M_s^2/M_A$ and $\propto l^{-3/4}M_s^2$ respectively for slow and fast modes. When $M_s$ is large we expect to have no spatial dependence $l$ in the power law . {\torefereeone From Table \ref{tab:pred} we see that (1) The Alfven modes do not introduce any density fluctuations, so we can neglect it in the time being. (2) The contribution of slow modes to $\nabla I$ is $\propto M_s$ (3) The contribution of fast modes to $\nabla I$ is $\propto M_A$ (4) The relative slow-to-fast mode energy ratio in our simulations with random driving at $\beta<1$ is roughly 3:1. With these factors taken into account, we can see that $M_A$ is inevitably contributing to both $\sigma_{\nabla I}$ and $\sigma_{\nabla C}$ depending on the weighting of fast modes.   Yet from the same argument, we see that the fast modes are subdominant in all of our simulations. In this scenario we expect that the slow mode dependence (i.e. $\propto M_s$) would dominate over that of fast modes.} The non-linear effect that $M_A$ has in the power-law might {\torefereeone affects} the prediction of the $M_s$ using Table \ref{tab:pred}. In view of that, we prepared the two subsets of data "m0-m9" and "huge-0-huge-9" which {\torefereeone carry} fairly close $M_s$ but a wide range of $M_A$ to see how would $M_A$ change the predictions in Table \ref{tab:pred}. We divide standard variations of $\nabla I'$ of these two sets of data and compared to the mean value of them and find that the standard deviation accounts for 14\% and 6\% only for the sets "m0-m9" and "huge-0-huge-9" respectively. {\torefereeone Thus we believe that $M_A$ would be a less important factor compared to $M_s$ as on the left of Fig. \ref{fig:n.IGVCGn}.}

We showed in \S \ref{subsec:stat_observation} that the centroid gradient amplitude is not the {ideal} variable in obtaining the sonic Mach number due to the composite product of density and velocity, making {us} difficult in isolating the effect of density from velocities {in centroids}. It would still be interesting to see how the centroid gradient amplitude would behave as a function of sonic Mach number. Here we would expect that the velocity centroid would follow a power law to be an undetermined mixture of density and velocity power law as predicted from Table \ref{tab:pred}. As we see from \S \ref{subsec:stat_incompressible} that the three dimensional velocity gradient amplitude relation (See e.g. Eq.\ref{eq:vga_fundamental}) does not carry a factor of $M_s$, meaning that the resultant centroid gradient amplitude term would solely coming from the density contribution.

In the computation of velocity centroid, we would simply compute $\sigma_{\nabla C} = \left(\langle \left(\nabla C\right)^2\rangle-\langle\nabla C\rangle^2\right)^{1/2}$ instead of diving the mean value of $C$ since velocity terms do not require such a normalization {\torefereeone as shown in \S \ref{sec:theory}}. On the right of Fig. \ref{fig:n.IGVCGn} we show the {\torefereeone variations} of $\sigma_{\nabla C}$ as a function of $M_s$. Due to the cumulative contribution of density and velocity terms, it is expected to have some power law that deviates from $M_s$ or $M_s^2$. We can see that while the blue points are generally following the $\propto M_s$ power law, the red data points {\torefereeone exhibit} a more scattered pattern. {\torefereeone Yet the centroid gradient amplitude does shown to suffer from less influences from the line of sight angle effects, which we will discuss in\S \ref{subsec:angles}).}

\section{Robustness of the methods}
\label{sec:robust}

\subsection{The angle between magnetic field and line of sight}
\label{subsec:angles}
In \S \ref{subsec:stat_observation} we expect that the angle {\torefereeone between the mean magnetic field and the line of sight} would also be an important factor in affecting the computed gradient amplitude. While the turbulent eddies are anisotropic following the theoretical prediction in \cite{GS95}, it is shown that \citep{2014ApJ...790..130B, 2018ApJ...865...54Y} that the correlation function anisotropy of velocity centroids drops significantly from anisotropic to merely isotropic as the relative angle $\theta$ between the mean magnetic field $\langle B \rangle$ and the line of sight directions decreases. {\torefereeone Readers should be careful that the study performed in \S \ref{sec:analysis} has the mean magnetic field being perpendicular to the line of sight, i.e. $\theta=90^o$. }.

\begin{figure*}[t]
\centering
\includegraphics[width=0.49\textwidth]{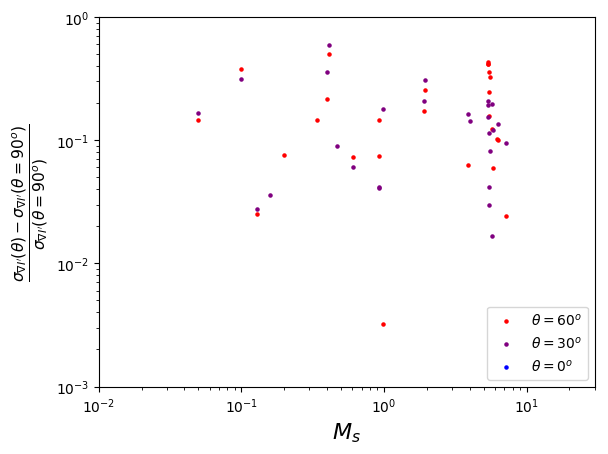}
\includegraphics[width=0.49\textwidth]{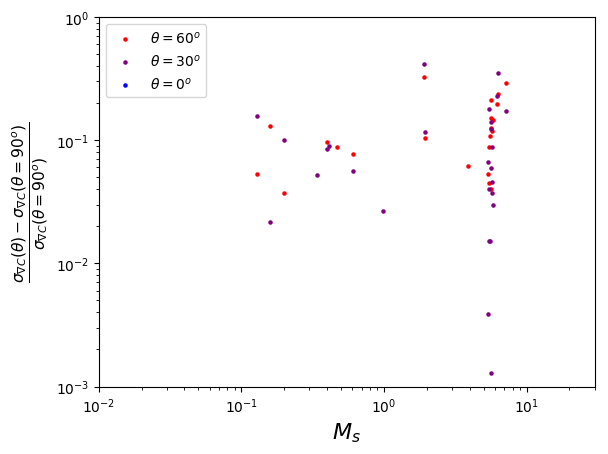}
\caption{\label{fig:n.angle} \torefereeone Two figures showing the fractional change (See Eq.\ref{eq:fc}) of both intensity gradient amplitude dispersion $\sigma_{\nabla I'}$ (left) and the centroid gradient amplitude dispersion $\sigma_{\nabla C}$ (right)  respond as functions of  the relative angle $\theta$ (in degrees) between the mean magnetic field $\langle B \rangle$ and the line of sight directions (In the case of Fig.\ref{fig:n.IGVCGn}, we have $\theta=90^o$) for different Ms. }
\end{figure*}

{\torefereeone 
To characterize how the relative angle $\theta$ would change the result in Fig.\ref{fig:n.IGVCGn}, we consider the following quantity that records the fractional change when we vary the $\theta$ :
\begin{equation}
    X = \frac{\sigma(\theta)-\sigma(\theta=90^o)}{\sigma(\theta=90^o)}
    \label{eq:fc}
\end{equation}
which evaluates the relative changes of the dispersion compared to the value we have in Fig.\ref{fig:n.IGVCGn}. If the quantity $X$ is small for $\nabla I'$ and $\nabla C$ and is not a function of $M_s$ (See.\ref{subsec:stat_observation}), then we expect that the shape of Fig.\ref{fig:n.IGVCGn} should kept unchanged. On the left and right of Fig \ref{fig:n.angle} shows how the fractional change is correlated to the relative angle $\theta$ for our simulations for intensity gradient amplitude dispersion $\sigma_{\nabla I'}$ and centroid gradient amplitude dispersion $\sigma_{\nabla C}$ respectively. We can see that the fractional change is generally in the order of 0.1 and has no visually recognizable trend as a function of $M_s$ With these fluctuations we reported that the power-law that we see in Fig.\ref{fig:n.IGVCGn} with $\theta=90^o$ is still seen for other choice $\theta$. From the argument above, we therefore conclude that the effect of $\theta$ to the power law we discussed in Fig.\ref{fig:n.IGVCGn} is not significant.
}

\subsection{Test on the noise sensitivity}

One of the biggest potential discrepancy of the gradient amplitude method is the noise. In the previous development of VGT, the noise is tackled by either a global Gaussian kernel or a low-pass Fourier filter (See \citealt{Letal17}). For the purpose of tackling the {\it orientation} of gradients these methods worked well and have been applied to observations (e.g. \citealt{survey,velac}). We here would like to test whether the same strategy works for the gradient amplitude method. 

\begin{figure}[t]
\centering
\includegraphics[width=0.49\textwidth]{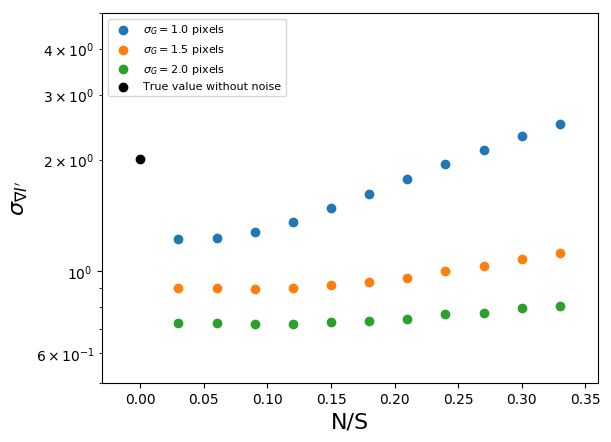}
\caption{\label{fig:n.noise}  A figure showing the response of $\sigma_{\nabla I'}$ with respect to the noise-to-signal level $N/S$ under different Gaussian smoothing kernels (in pixels).}
\end{figure}

We pick the intensity map of the cube "h0-1600" as an example. We gradually add a Gaussian-model white noise to the intensity map with the strength set to be some factor relative to the dispersion of the intensity map. The strength of the noise compared to the dispersion of the intensity map is simply the noise-to-signal level ($N/S$). Since a general practice in observation is to exempt data that has $S/N<3$, therefore we would only keep adding white noise up to $N/S<1/3$. We would then smooth with Gaussian filters with different width $\sigma_G$ in units of pixels. Fig. \ref{fig:n.noise} shows how the $\sigma_{\nabla I'}$ responds to the noise levels as we change the size of the Gaussian filter.  We could see  that when $\sigma_G$ arrives 4 pixels , there is simply no recognizable fluctuations for $\sigma_{\nabla I'}$  as $N/S$ increases. We therefore believe that noise would not be a concern for the intensity gradient amplitude method.

\section{Determining the stage of collapse using the gradient amplitude statistics}
\label{sec:gravity}
It has been suggested that in the stage of gravitational collapse gradients of intensities and centroids would turn from being perpendicular to magnetic field to parallel {\it gradually} \citep{YL17b, LY18a} which was termed "re-rotation". The concept of re-rotation was applied to recent analysis of observations \cite{survey,velac}. However the practical procedure of re-rotation still requires further studies. Here we have to be cautious that we are not discussing the statistics of gradient relative orientation to magnetic field (\citealt{Soler2013}, see also \citealt{IGVHRO} for a comparison) as a function of column density, but by measuring the differences of the velocity gradient orientation in diffuse media and self-gravitating cores as a means of determining the direction of magnetic field without using polarimetry.

Aside from the change of orientations, we should also expect {\torefereeone changes} of the amplitude of gradients in different stages of gravitational collapse. It is natural that the amplitude of gradients for both densities and velocities would increase due to the acceleration of gravitational field. Yet, such signature could be possibly confused by strong compressions from shocks. MHD theory tells us that the maximal density enhancement for shock is bounded by either the sonic or the Alfvenic Mach number, depending on the plasma $\beta$\citep{2019ApJ...878..157X}. Comparatively, the density enhancement from gravitational collapse is unbounded. When we compute the amplitude of gradient with respect to the ambient mean value, one could possibly see an exponential growth of the gradient amplitude of observables in the self-gravitating region but not the case for shocks. By using numerical simulations to enumerate the collapse of interstellar media, we could test how gradient amplitude would aid determining the stage of collapse. To investigate the effects of gravity we evolve the numerical cube $h0$ to the stage until the high density structure is not resolvable, which is named Truelove-Jeans criterion \citep{1997ApJ...489L.179T}.

We use two frequently used physical parameters to characterize the stage of the collapse. The first one is the average gravitational energy density $\epsilon_{ge} = \langle -\rho \Phi\rangle$, which is a simple estimate of how clumpy the density structure is when responding to the gravitational field. Notice that in our periodic simulations $\langle \Phi\rangle=0$ and $\langle \rho \rangle =\text{const}$ guaranteed in all times.  The other parameter that is related to the stage of collapse is the relative free fall time $t_{rff}$ which is a measure of how long the system has been collapsed since gravity is switch on. We compute the free-fall time for those pixels having the gravitational potential being negative and compare to the free-fall time assuming the whole system is going to collapse, i.e. $t_{ff,0}\sim(G\langle \rho \rangle)^{-1/2}$. 
\begin{equation}
\begin{aligned}
    t_{rff} &= t_{ff,0} - t_{ff}\\
    &\sim (G\langle \rho \rangle)^{-1/2} - (G\langle \rho \rangle_{\Phi<0})^{-1/2}
\end{aligned}
\end{equation}
The difference between them indicates the time that the collapse has taken. From our discussion, we expect that the part of the region that has gravitational collapse will have a higher density value, while the ambient environment would have a lower one, which apparently suggests that we should use the dispersion of gradient amplitudes to measure the effect of gravity. However that could not be used directly because the enhancement of density by gravity is also a function of density. We therefore would compute the parameter $\sigma/\mu$ which would normalize the dispersion quantity, hopefully reduce the density contribution from the growth of gradient amplitude.

\begin{figure*}[t]
\centering
\includegraphics[width=0.49\textwidth]{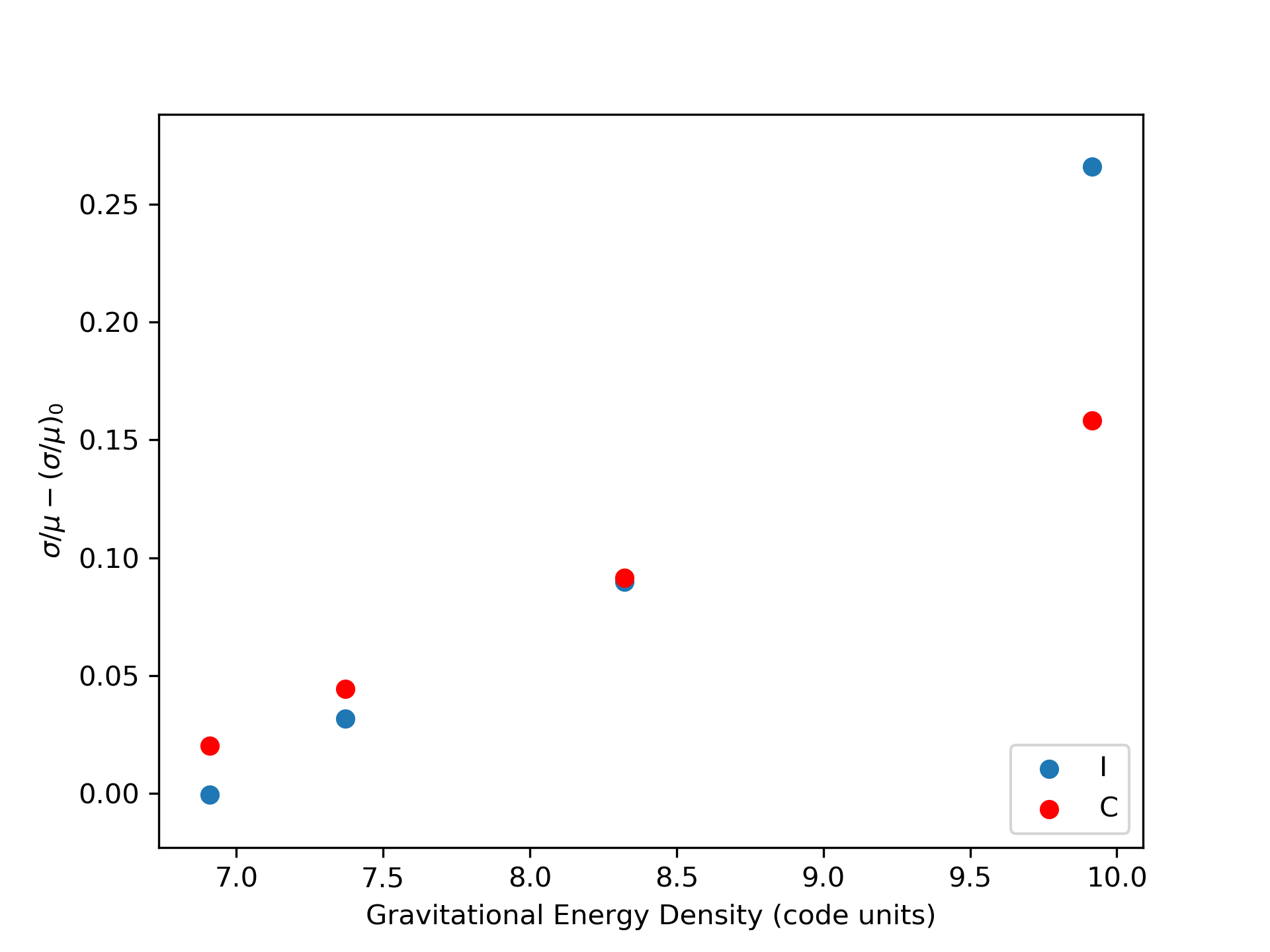}
\includegraphics[width=0.49\textwidth]{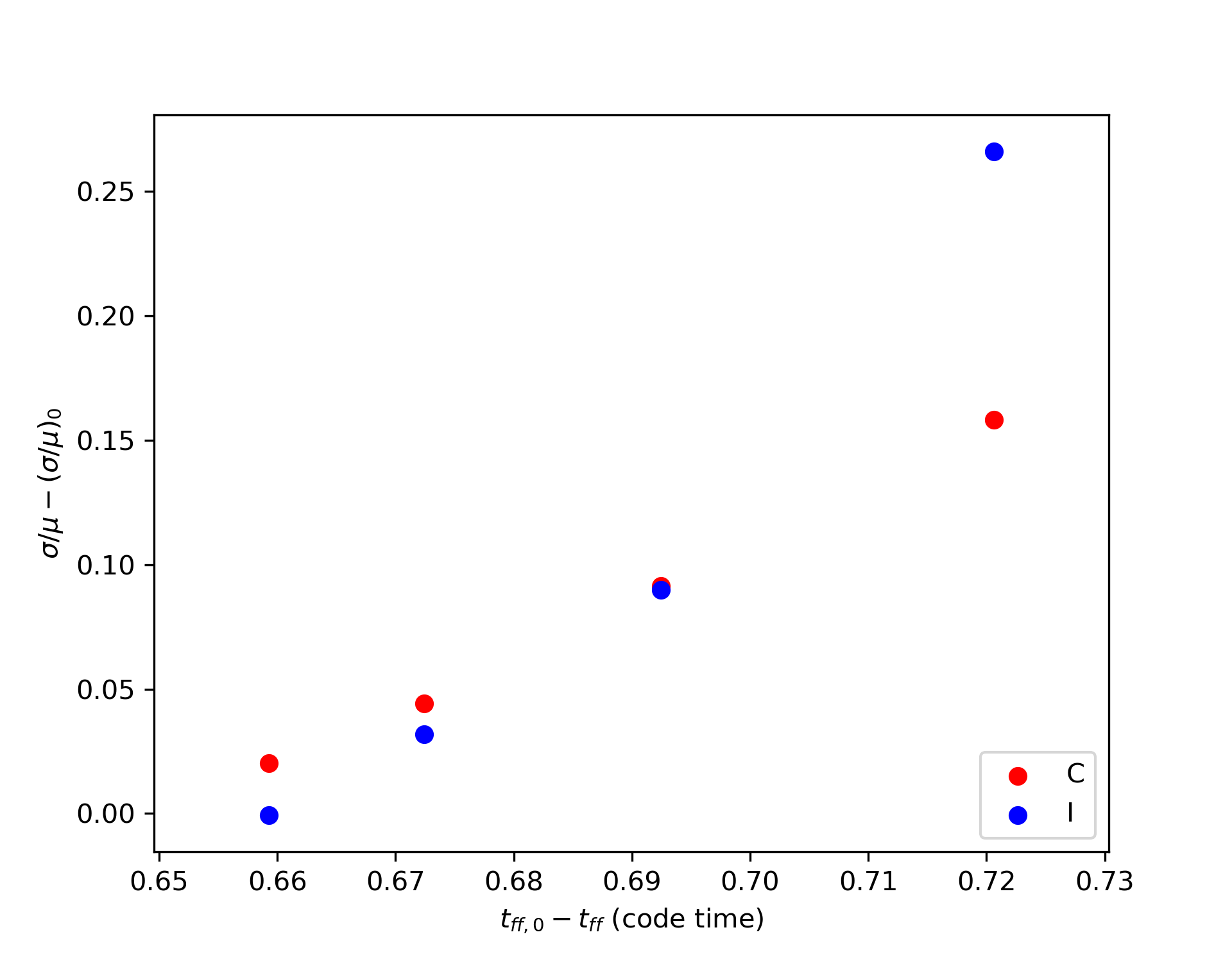}
\caption{\label{fig:n.grav} \torefereeone Two figures showing the response of $\sigma/\mu$ with two commonly used physical parameters measuring the stage of collapse: $\sigma/\mu$ as a function of the gravitational energy density (Left) and  relative free fall time (Right) for both intensity and centroid.}
\end{figure*}

Fig. \ref{fig:n.grav} shows the response of $\sigma/\mu$ with respect to $\epsilon_{ge}$ and $t_{rff}$ which we include also the centroid gradient amplitude statistics. We see that there is a linear relationship between $\sigma/\mu$ with respect to $\nabla I'$. We fit the linear relation and see that
\begin{equation}
\begin{aligned}
\left(\frac{\sigma}{\mu}\right)_{I} &\approx -0.62 + 0.09 \epsilon_{ge}\\
\left(\frac{\sigma}{\mu}\right)_{I} &\approx -2.89 + 4.35 t_{rff}\\
\end{aligned}
\end{equation}
with $R^2>0.9$ which gives some insight of the stage of the gravitational collapse when one investigates the gradient amplitude statistics in self-gravitating media. For instance, \cite{survey} provides a number of clouds with a strong signature of gravitational collapse as indicated by VGT. With our technique we can investigate {\it quantitatively how strong} the gravitational collapse in terms of free-fall timescales or gravitational energy density, which can possibly provide the probability of star formation in a given region.

\section{Deconvolution of velocity channel map}
\label{sec:deconvolution}

The sonic number not only could possibly help observers to characterize the physical conditions of turbulence in molecular cloud but also allows observers to extract the statistics of turbulence in velocity channel maps {\torefereeone (See \S \ref{subsec:caveats} for the discussions of the strength and caveats of our method)}. In fact, velocity channels are one of the main observables in the series of VGT papers (\citealt{YL17b,LY18a,LYH18}) and their gradients have shown to be better in tracing magnetic field directions compared to the gradients of intensity and centroid maps both numerically (e.g. \citealt{LY18a}) and observationally (e.g. \citealt{survey,velac}). Aside from magnetic field tracing, techniques such as Velocity Channel Analysis (VCA) and Velocity Coordinate Spectrum (VCS, \citealt{LP00}, see also their work on \citeyear{LP04,LP06,LP08}) also use the statistics of velocity channel maps to predict the three dimensional velocity and density spectrum from observation. 

\cite{LP00} suggests that, in the absence of thermal broadening, when the velocity channel is thin, i.e. the velocity window width of the velocity channel is smaller than the characteristic velocity in the velocity channel, then  the fluctuation of velocity channels are dominated by velocity fluctuations if the 3D density spectrum is steep ($n_\rho<-3$). In finite temperature $T$, the effect of velocity dominance in velocity channel map is called "velocity crowding". The velocity crowding effect modifies intensity enhancements in channel maps and also flattens the power spectrum of velocity channels.

Since we have shown in \S \ref{sec:analysis} that intensity and centroid gradient amplitude maps have the similar power-law dependencies to $M_s$ is predicted under the \cite{GS95} framework, it is very natural to imagine the same argument as in \S \ref{sec:theory} would apply also to velocity channel map. However in the presence of strong thermal broadening ($M_s\ll1$), the channel map is effectively a weighted integral of the intensity map and thus the velocity channel map gradient amplitude in such limit would have the same $M_s$ dependence as their integrated intensity counterpart even through \cite{LP00} already formulated analytically how thermal broadening would alter the channel map. The thermal broadening is important for light species, especially for HI and H$_\alpha$ lines. For them the corresponding masking of velocity crowding effect via thermal broadening prevents the proper use of VCA and VCS techniques. 

A workaround for such a problem would be to remove the thermal broadening effect from the velocity channel maps. Such method is possible since thermal broadening is simply a Gaussian convolution with $c_s$ as the broadening width. The estimation of the $c_s$ is not easy. However since the thin channels and the thermally broadened thin channels follow statistics derived by \cite{LP00}, it is possible for us to recognize such a transition by performing Wiener deconvolution algorithm through a list of trial $c_s$ even in the presence of noise. We shall see in \S \ref{subsec:mathppv} how such mathematical constructions would allow us to retrieve $c_s$. In the isothermal limit, if we have an estimation of $M_s$ as listed in \S  \ref{sec:analysis}, with the measured spectral line width one could possibly acquire also the injection velocity $v_L= M_s c_s$, which is one of the central physical quantities determining the properties of turbulence. In \S \ref{subsec:mathppv} we would discuss the necessarily mathematical foundations of PPV thermal broadening and the method of deconvolution. In \S \ref{subsec:cs} we discuss how to obtain $c_s$ by a self-consistent algorithm and in \S \ref{subsec:numppv} we illustrate how the measured $c_s$ could allow one to reverse engineer the PPV cube without thermal contributions.

\subsection{Mathematical formulation on thermal broadening and deconvolution}
\label{subsec:mathppv}
Mathematically, the density in PPV space of emitters with local sonic speed $c_s(\bf x) = \sqrt{\gamma k_B T/\mu_{MMW}}$, where $\mu_{MMW}$ is the mean molecular weight of the emitter, moving along the line-of-sight with stochastic turbulent velocity $u(\bf x)$ and regular coherent velocity, e.g. the galactic shear velocity, $v_{\mathrm{g}}(\bf x)$ is \citep{LP04}
\begin{equation}
\rho_s(\mathbf{X},v) =\int_0^S \!\!\! dz 
\frac{\rho(\mathbf{x})}{\sqrt{2\pi \beta_T}}
\exp\left[-\frac{(v-v_{\mathrm{g}}(\mathbf{x}) -
u({\bf x}))^2} {2 c_s({\bf X} ,z)^2 }\right] 
\label{eq:rho_PPV}
\end{equation}
where sky position is described by 2D vector $\mathbf{X}=(x,y)$ and $z$ is the line-of-sight coordinate, $\gamma$ is the adiabatic index. Notice that $c_s$ would be a function of distance if the emitter is not isothermal. The Eq.~(\ref{eq:rho_PPV}) is \textit{exact}, including the case when the temperature of emitters varies in space. The observed velocity channel at velocity position $v_0$ and channel width $\Delta v$ is then, assuming a constant velocity window $W(v)=1$,
\begin{equation}
\begin{aligned}
Ch(\mathbf{X};v_0,\Delta v) 
= &\int_{v_0-\Delta v/2}^{v_0+\Delta v/2}dv \rho_s(\mathbf{X},v)\\
= \int_0^S dz \frac{\rho(\mathbf{x})}{\sqrt{2\pi c_s^2}} &\int_{v_0-\Delta v/2}^{v_0+\Delta v/2} dv e^{-\frac{(v-v_{\mathrm{g}}(\mathbf{x}) - u({\bf x}))^2} {2 c_s^2 }}
\end{aligned}
\label{eq:channel}
\end{equation}

The Eq.~(\ref{eq:rho_PPV}) represents the effect of the velocity-dependent mapping from the three-dimensional Position-Position-Position (PPP) space to PPV space.  Due to this mapping, the PPV density $\rho_s(\mathbf{X},v)$ at a given velocity $v$ is determined both by the spatial density of the emitters $\rho(x,y,z)$ and their respective line-of-sight velocities.  Note that formal caustics, understood as singularities of differentiable map from PPP to PPV space, arise only in the limit of $c_s\to 0$.  

The Eq. \ref{eq:rho_PPV} can be inverted if $c_s$ is known. We can then connect the thermally broadened position-position-velocity cube  density $\rho_s({\bf X},v)$ to that of the underlying PPV density $\rho_v({\bf X},v)$:
\begin{equation}
    \rho_s({\bf X},v) = \int dv' \rho_v({\bf X},v') e^{-\frac{(v-v')^2}{2c_s^2}}
    \label{eq:devolve}
\end{equation}
which is a convolution of the raw PPV density $\rho_v({\bf X},v)$ with respect to the thermal Gaussian kernel $G(v) = \frac{1}{\sqrt{2\pi c_s^2}}\exp(-\frac{v^2}{2c_s^2({\bf X})})$, a function of the sonic speed. 

\subsection{Mapping $c_s$ by constructing exponential reduced centroids}
\label{subsec:cs}
We consider the following integral
\begin{equation}
\begin{aligned}
    \rho_e({\bf X},v) 
    = &\frac{\int_{|v|>\sqrt{2}c_s'} dv' \rho_s({\bf X},v')e^{{\text{\color{red} +}}\frac{(v-v')^2}{2c_s^2}} }{\int_{|v|>\sqrt{2}c_s'}e^{{\text{\color{red} +}}\frac{(v-v')^2}{2c_s^2}}}
    \label{eq:erc}
\end{aligned}
\end{equation}
which integrates the observed PPV density $\rho_s({\bf X},v')$ with an exponential factor that looks like {\it the inverse convolution} of the Gaussian kernel $G(v)$ which we term the exponential reduced centroid (ERC). However we must remind the reader that the expression in Eq.\ref{eq:erc} is not a deconvolution of the thermal function.

\begin{figure*}[t]
\centering
\includegraphics[width=0.48\textwidth]{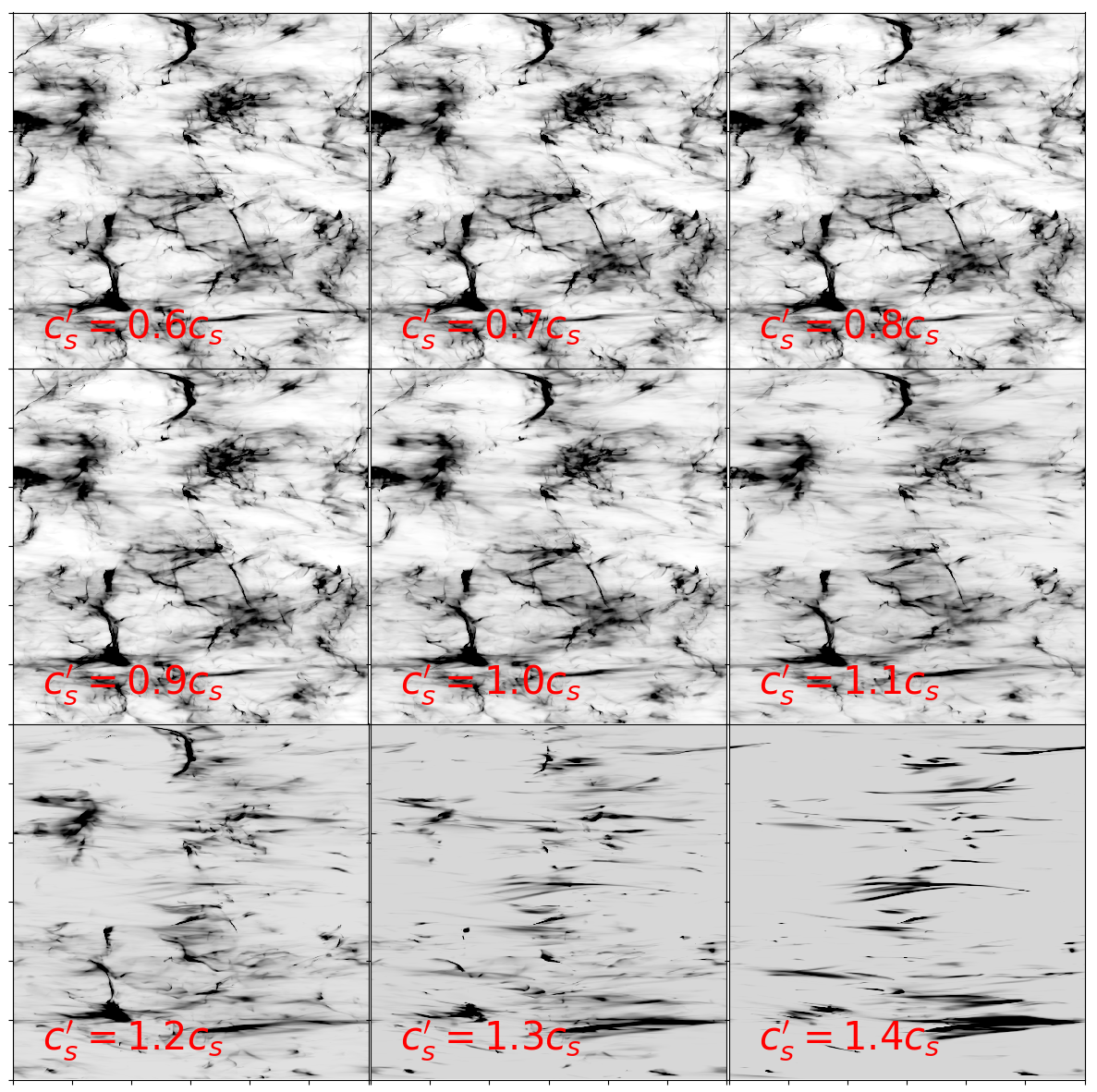}
\includegraphics[width=0.48\textwidth]{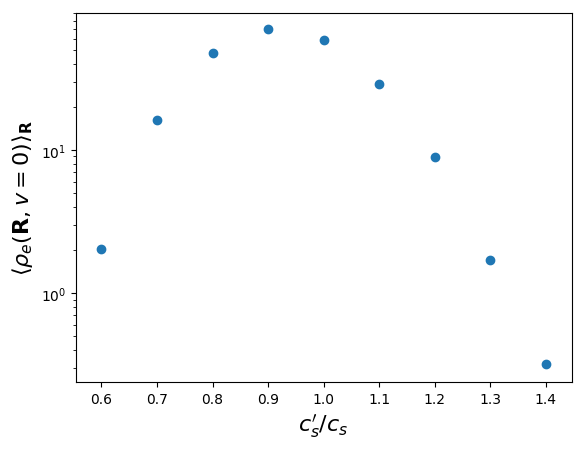}
\caption{\label{fig:n.erc}  (Left) A set of illustrative figures showing how the exponential reduced centroid (ERC) would behave as a function of prior $c_s'$. The colorbar scale is relative for all maps and set similarly as Fig.\ref{fig:illus}. (Right) The mean amplitude of the ERC as a function of $c_s'/c_s$. }
\end{figure*}

\begin{figure}[t]
\centering
\includegraphics[width=0.48\textwidth]{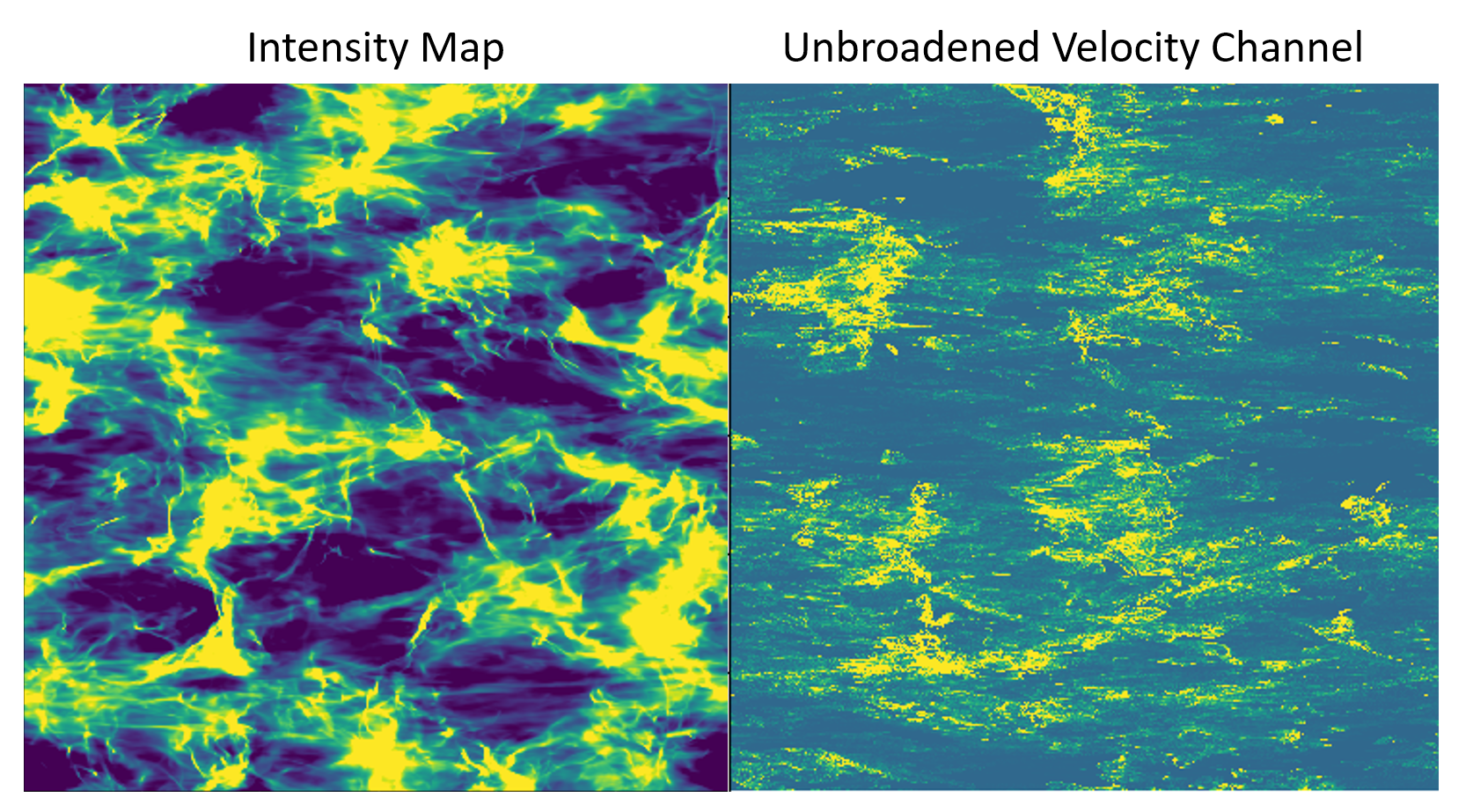}
\caption{\label{fig:n.I}  The intensity and the velocity channel of the cube h0-1200 when we set $c_s = 10c_{s,actual}$. }
\end{figure}

To perform the study numerically, we prepared a series of synthetic PPV density by using the PPV cube of h0-1200 and convolve the simulations with different values $c_s$ from $0.25 c_{s,actual}-10.0 c_{s,actual}$, where $c_{s,actual}$ is the actual value we put into the simulation. We then compute the ERC function following Eq.\ref{eq:erc}. We show pictorially how does the ERC behave on the left of Fig. \ref{fig:n.erc} as we vary our guess of $c_s'$. When $c_s'$ is small, the ERC is actually intensity-like (Readers could compare the results from Fig.\ref{fig:n.erc} to that in Fig. \ref{fig:n.I}), meaning the cross correlation between the ERC and intensity map is high. When $c_s'$ is large, the ERC becomes velocity like and has obvious striations that is not from the density structures. On the right of Fig. \ref{fig:n.erc} shows how the mean amplitude of the ERC $\rho_e({\bf X},v')$ behaves as a function of $c_s'/c_s$. We see that the mean amplitude of the ERC follows a special property that it peaks at around $c_s'/c_s\sim 1$ (Exact value: 0.9). We see this property around different setting on $c_s'$ and $c_s$ and we believe this provides a unique way in probing the sonic speed.

\subsection{Wiener Deconvolution Algorithm}
\label{subsec:numppv}
Since we already obtain an estimate of $c_s$ from the previous subsection, we can then thus deconvolve the raw PPV density $\rho_v({\bf X},v)$ from the observed PPV cube $\rho_s({\bf X},v)$. Assuming we have a prediction of sonic speed $c_s'$ (which might be different from true sonic speed $c_s$), we see that the Wiener deconvolution algorithm would allow us to estimate of underlying PPV density $\rho_v'$ if the noise-to-signal ratio is provided.
\begin{equation}
    \mathcal{F}\{\rho_v'\} = \frac{\mathcal{F}\{G(c_s')\}\ \cdot \mathcal{F} \{\rho_s\}}{|\mathcal{F}\{G(c_s')\}|^2 + (N/S)^2}
\end{equation}
where $\mathcal{F}$ is the Fourier transform operator and  $N/S$ is the noise-to-signal ratio. In other words, the actual primitive PPV density $\rho_v$ and the estimated PPV density $\rho_v$ would be related by:
\begin{equation}
    \mathcal{F}\{\rho_v'\} = \mathcal{F}\{\rho_v\}\frac{2\pi c_s c_s' e^{-2\pi k^2 (c_s^2+c_s'^2)}}{e^{-4\pi k^2 c_s'^2}+(N/S)^2}
    \label{eq:rho_vv}
\end{equation}

\begin{figure*}[t]
\centering
\includegraphics[width=0.96\textwidth]{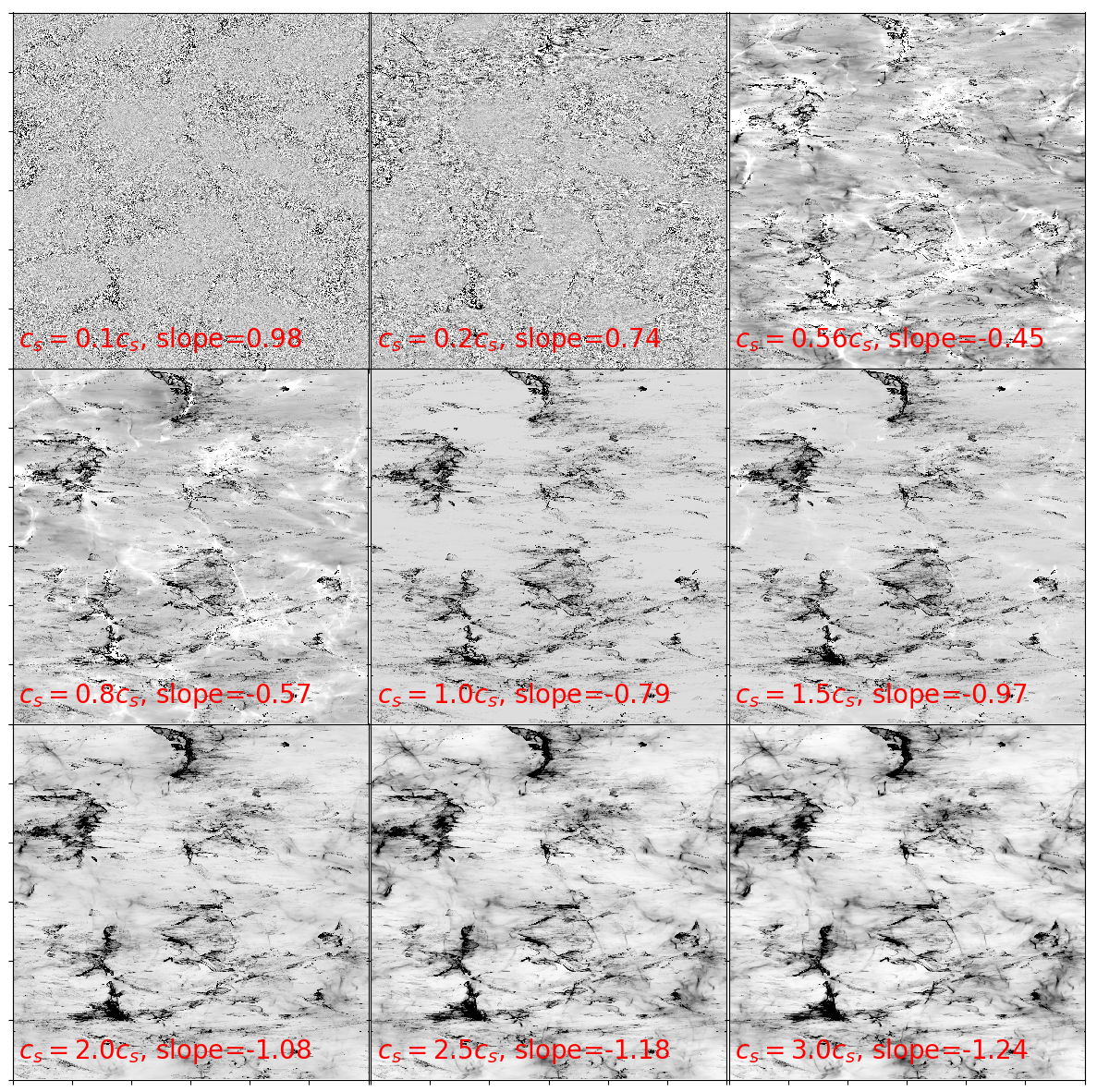}
\caption{\label{fig:n.WDA} A figure showing the deconvolved PPV cube at $v=0$, i.e .$\rho_v({\bf R},v=0)$ as a function of $c_s'/c_s$, with the label including the spectral slopes. }
\end{figure*}

Figure.\ref{fig:n.WDA} shows how the algorithm deconvolves the PPV cube in the presence of noise and the variation of $c_s$. One could see that when $c_s'<c_s$, the map is basically noise-like and has a power spectrum with positive slope. When $c_s'>c_s$, the structure of the velocity channel map would be similar to that we show on the left of Fig.\ref{fig:n.I}. Only when we choose an appropriate $c_s'\sim c_s$ which could be recognized by spectral slopes, one could see velocity like structures that what we displayed on the right of Fig. \ref{fig:n.I}.

\section{Discussions}
\label{sec:discussion}
\subsection{Importance, possible future studies and caveats of our work}
\label{subsec:caveats}
Our work here focus on obtaining two of the most important astrophysical quantity based on the amplitude statistics of the observables, namely the sonic Mach number (\S \ref{sec:theory},\ref{sec:analysis}) and the sonic speed (\S \ref{sec:deconvolution}). We start from theory that there should be an expected correlation between the observed intensity gradient amplitude and the underlying sonic Mach number, which we also confirm numerically. As a separate development, the sonic speed is obtained by probing the mean amplitude of the exponential reduced centroid, which allows us to not only estimate the injection velocity $v_L = M_s c_s$ but also the true, thermally unbroadened position-position-velocity cube (\S \ref{subsec:numppv}). The power of amplitude statistics with the support of the theory of MHD turbulence \cite{GS95,LV99,CL03} and observational diagnostics \cite{LP00,LP04,LP06} provides a unique way in probing the physical conditions and extract turbulence statistics in observed spectroscopic data. Moreover, with the PPV cube without thermal broadening, we can safely use the tools developed from \cite{LP00,LP04} about the statistics of turbulence both parallel and perpendicular to the line of sight. This shows that the amplitude statistics has its unique position in predicting the turbulence statistics, and could possibly be used in situations that has more physics involved, e.g. multiphase media, gravito-magnetohydrodynamics.

One of the small but very important aspect is the constant dependence of velocity gradient amplitudes to both sonic and Alfvenic Mach number as when $M_s\gg 1$. This constancy is expected from theory but currently we cannot possibly apply it to observations since we do not observe the pure velocity map from spectroscopic data. This should be further studied both theoretically and numerically. The theory that we derived in \S \ref{sec:theory} depends also on the mode composition in the astrophysical turbulence, which is also a quantity that has no easy option to obtain. While there are attempts (\citealt{2018arXiv180801913Z,2019arXiv190701853M} ,Chepurnov et.al 2020) in acquiring them both numerically and based on the observed synchrotron data, such a method is not extended to the regime that we are tackling now. Due to the differences of the dependence of the $M_s$ and $M_A$ for the three different modes, it is possible to use gradient amplitude as a quantifier to estimate the relative weights of the modes, or at least the ratio between compressible and incompressible modes. We could see from Table \ref{tab:pred} that, in the case of sub-Alfvenic systems, there is a dependence on $M_A$ for Alfven modes but not for the two compressible modes. Measuring the value of $M_A$ could probably allow us to estimate the ratio of the modes since the variance of the gradient amplitudes from different modes is contributed by the squared sum of their respective dispersions. By modelling this the ratio of modes should be able to obtain.

One of the very important caveat here that requires further study is the existence of rotation curve. In our discussion in \S \ref{sec:deconvolution} we do not take the effect of galactic shear into account since we would like to simplify our deviation. However for applications to large-scale data like the HI-maps, it is necessary to also consider the effect of galactic rotation. Fortunately this could be possibly studied both numerically and observationally by introducing a proper rotation curve model that corresponds to our galaxy. The quality of the wing channel is also a concern in the method of \S \ref{sec:deconvolution}, since if there is only a limited number of the channels provided, the deconvolution algorithm might not have enough data in acquiring the PPV cube. {\torefereeone Another caveat in the gradient amplitude is that, there do exist a variation of intensity gradient amplitude statistics due to the different weighting of modes and the line of sight effects, which both of them are hard to estimate in observations. Comparatively, the centroid gradient amplitude would be more robust in terms of real application if one could accept $\sigma_{\nabla C}\sim M_s^\alpha$ with $1<\alpha<2$ as acquired in Fig \ref{fig:n.IGVCGn}. }

\subsection{Importance to the development of MHD turbulence theory and interactions between theory and observations}

We discussed in \S \ref{sec:theory} that the whole idea of applying GS95 to real life astrophysical observations is to have some sort of statistical averaging on some MHD variables, with proper estimations on the observational effects. For example, one should consider the effect of projections if the observable is obtained by some other variables collected along line of sight. If a velocity axis exists along the line of sight then it is important to understand how the velocity effect contributes to the observable. The statistical averaging, through implicitly applied, could actually be found in a vast number of literature. For instance, the use of spectra \citep{MG01}, velocity-distance relation \citep{1981MNRAS.194..809L}, correlation and structure functions \citep{EL05}, histogram of orientations \citep{Soler2013} and block averaging \citep{YL17a}. The current work uses the same principle as the same author did three years ago in their work \cite{YL17a}. The same idea has been brought to here by simply using the mean (\S \ref{subsec:mathppv}) and the dispersion (\S \ref{subsec:stat_observation} of the observables. It is expected that the theory could predict with observations better by testing which statistical quantifier would be the best in extracting the statistical behavior from the observables based on the GS95 scaling. 

\subsection{Implications to the development of the Velocity Gradient Technique}

The Velocity Gradient Technique in its current form provides a reliable way of tracing magnetic field directions. Our work here is the first attempt to connect the {\it gradient amplitudes} to $M_s$ and quantify their relation through the same principle that formulates VGT. We note that the visual correspondence that we demonstrated is also handy for observers to approximate the physical conditions in the parts of the ISM. The prospects of using gradients to probe physical conditions further increases the value of the Velocity Gradient Technique by  helping to further constrain the physical conditions in different ISM phases. We expect that this should help in choosing between different models of star formation (See \citealt{C10,2012ARAA...50..29,LC12}).

This work provides the values of $M_s$ in a cost-effective way and complementary to other papers on VGT targeting Mach numbers, e.g. \cite{LYH18} uses dispersion of velocity gradients to estimate $M_A$. Synergistic use of these methods provides a way to cross-check the measurements. Moreover, in case when polarimetry data is unavailable, the gradient technique provides an alternative way to study the physical conditions of interstellar media.

\subsection{Role of thermal broadening}

The VGT has become a sophisticated technique applicable to studying both subsonic and supersonic environments. For probing VGT for subsonice turbulence in practice one can use different approaches. First of all, the velocity centroids are not contaminated by thermal broadening (see Esquivel \& Lazarian 2005, Kandel et al. 2016) and therefore the centroids gradients can be used if the thermal broadening exceeds the turbulent one. Reduced centroids (Lazarian \& Yuen 2018) can be applied to the spectral line data broadened by galactic rotation. 

Channel maps provide a valuable way of analysing the data. In the case of HI, it was noted in Lazarian \& Pogsyan (2000) that while the Warm HI may dominate at high galactic latitudes in terms of total emissivity, the contribution from cold HI is expected still to dominate in thin channels. Therefore, if the current view of the two phase turbulent HI, namely clumps of Cold HI moved together and by the Warm HI is true, the statistics of thin channels represents the statistics of velocity with cold HI serving as a tracer of the Warm HI dynamics. This is explained in more details in Yuen et al. (2019).

This paper, however, discusses a way to deal with the thermal broadening through the deconvolution. This approach can be useful for, e.g. H$\alpha$ emission lines. While Lazarian \& Pogosyan (2000) showed that the thermal line width act as broad channels, the effect of thermal broadening is pretty simple in terms of its statistics. This allows the procedure of deconvolution that we demonstrated in this paper.

\subsection{Applying the results to various spectral lines}

Our study follows the empirical approach to obtaining the sonic Mach number that was explored in earlier papers (see \citealt{BL12}). The difference is that we use amplitudes of the gradients of velocity centroids. Our technique does not depend on the interpretation of 21 cm intensity enhancements in thin channel maps that has been debated recently (see \citealt{susan19, 2019arXiv190403173Y}). Moreover, HI is only one type of media to which the application of the technique is sought. Other optically thin lines, e.g. C18O, can be used. In analogy with the earlier studies exploring the effects of optical depth on the gradient technique (\citealt{2019ApJ...873...16H,velac,GLB17}), we expect that our present results can be applicable to CO lines with a moderate amount of self-absorption, e.g. 13CO lines. The corresponding study of different lines will be provided elsewhere.

The additional information obtained by the gradient amplitudes method is handy within the gradient technique. The accuracy of gradient technique in tracing magnetic field depends on the sonic Alfven Mach number $M_s$. In particular, the density enhancements associated with shocks can affect the intensity gradients. The increase of the amplitude of the gradients can help identifying shocks. The marginal dependence of the gradient amplitudes on the media magnetization and angle between the magnetic field and line of sight makes the present technique of studying $M_S$ rather robust. This testifies to the importance of the our suggested technique of identifying $M_s$. 

\section{Conclusions}
\label{sec:conclusion}
In this work, we explore how the amplitude statistics would help exploring the properties of interstellar turbulence through two aspects: the gradient amplitudes and the removal of thermal broadening. To summarize
\begin{enumerate}
    \item Based on the theory of MHD turbulence(\S \ref{sec:theory}), we derive a set of formulae for the amplitudes of three dimensional velocities (\S \ref{subsec:stat_incompressible},\ref{subsec:stat_compressible}) and densities (\S \ref{subsec:stat_dens})  for their dependencies to sonic and Alfvenic mach number as well as the dependencies to modes (Table \ref{tab:pred}). 
    \item We also derive similar predictions in gradient amplitudes of observables and discuss how to use them in observations (\S \ref{subsec:stat_observation})
    \item We tested our prediction in MHD simulations with simple statistical parameters. (\S \ref{sec:analysis}).
    \item We show that our prediction is robust to noise as long as a suitable gaussian kernel is used. (\S \ref{sec:robust})
    \item We discuss the potential use of the amplitude method in self-gravitating system. (\S \ref{sec:gravity})
    \item We discuss how the adoption of the amplitude method can help in deconvolving the observed velocity channel with strong thermal contamination (\S \ref{sec:deconvolution}).
\end{enumerate}

{\bf Acknowledgment.} We acknowledge Yue Hu and Ka Wai Ho for the fruitful discussions, and hospitality of the University of California, Santa Barbara in the Summer of  2019. We acknowledge the support the NSF AST 1816234 and NASA TCAN 144AAG1967.  This research used resources of the National Energy Research Scientific Computing Center (NERSC), a U.S. Department of Energy Office of Science User Facility operated under Contract No. DE-AC02-05CH11231, as allocated by TCAN 144AAG1967.

\appendix
\section{Extending to non-isothermal systems}
{\torefereeone

\begin{table}
\centering
\begin{tabular}{c c c c c}
Model & $M_S$ & $M_A$ & $\beta=2M_A^2/M_S^2$ & Resolution \\ \hline \hline
adiabatic $\gamma=1.00$ & 0.50  & 0.50 & 2.00 & $640^3$ \\
adiabatic $\gamma=1.67$ & 0.50  & 0.50 & 2.00 & $640^3$ \\
adiabatic $\gamma=2.33$ & 0.50  & 0.50 & 2.00 & $640^3$ \\ \hline \hline
\end{tabular}
\caption{\label{tab:sim2} Description of adiabatic MHD simulation cubes {which some of them have been used in the series of papers about VGT \citep{YL17a,YL17b,LY18a,2018ApJ...865...59L}}.  $M_s$ and $M_A$ are the R.M.S values at each the snapshots are taken. }
\end{table}

The aforementioned analysis focuses on the relation of the gradient amplitude to the sonic Mach number {\it in the case of isothermal environment}. For the isothermal simulations we employed in the main text, since the computation does not involve a definite length scale (e.g. involvement of gravitational length scale), therefore the only parameters that determine the saturated simulations are $M_s,M_A$ and the box length scale $L$. These simulations are scale free and can be rescale to other units to compare with observations.

However, in the case of diffuse interstellar media, where a distinct equation of state plus gas heating and cooling become important in understanding the dynamics and structure formations within ISM, we should expect the amplitude of gradients behave differently from the isothermal counterpart. For instance, in the case of adiabatic conditions for HI, where $P\propto \rho^{5/3}$ ({\torefereetwo See \citealt{2017NJPh...19f5003K,2018PhRvL.121b1104K}). In the case of phase equilibrium, the Warm and Cold neutral media can both be modelled by an adiabatic equation of state, see also \citealt{1995ApJ...443..152W} and reference therein}, the pressure exerted due to the crowding of density is larger than its counterpart in the isothermal system. As a result the gas tends to have a smaller density (or Intensity, in observation) gradient amplitude.  While it is true that the isothermal condition is generally true for giant molecular clouds ,such assumption is not correct for HI. Therefore there is a need for us to understand how the theory of gradient amplitude changes in the case of non-isothermal environment.

We test the result of gradient amplitude using the adiabatic numerical simulations in Table \ref{tab:sim2} and plot the $\sigma/\mu$ against polytropic index $\gamma$ in Fig \ref{fig:polit}. We see that when $\gamma$ increases $\sigma/\mu$ drops significantly. 
\begin{figure}[t]
\centering
\includegraphics[width=0.49\textwidth]{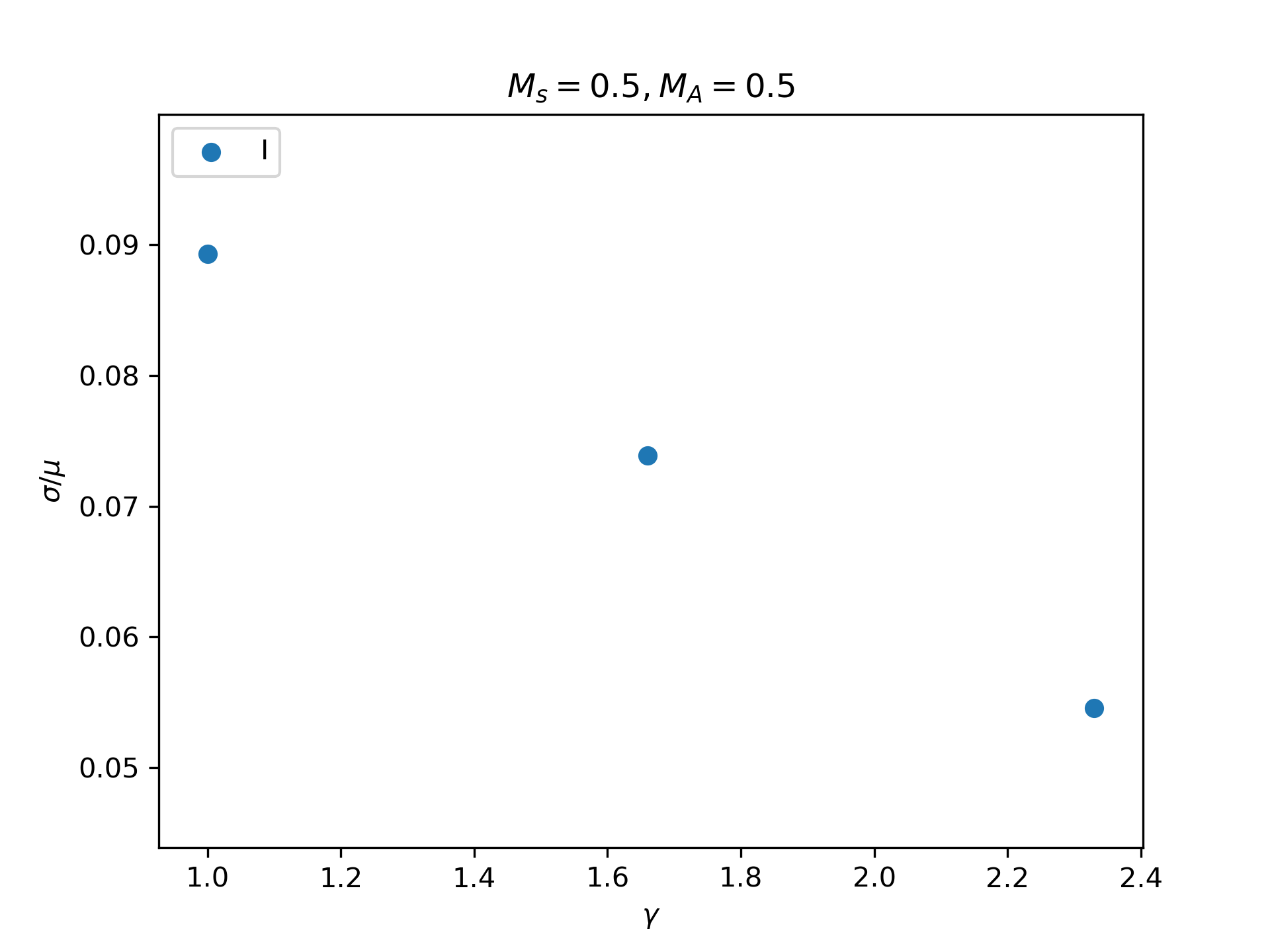}
\caption{\label{fig:polit} \torefereeone A figure showing the change of $\sigma/\mu$ with respect to the polytropic index $\gamma$ in adiabatic systems with the polytropic equation of state $P\propto \rho^{\gamma}$ and $M_s,M_A=0.5$}
\end{figure}

}

\end{document}